\newif\ifpdflatex    
\pdflatextrue           

\documentclass[fleqn,usenatbib,useAMS,twocolumn,numberedappendix]{aastex}
\usepackage{amsmath,esint}
\usepackage{url,aasmacros}
\usepackage{natbib}
\usepackage{soul, CJK}
\usepackage{multirow}
\usepackage{tablefootnote}
\usepackage{appendix}
\usepackage{bm}
\usepackage{xspace}
\usepackage{color}
\usepackage{enumitem}
\usepackage{booktabs}

\usepackage{graphicx}	
\usepackage{amsmath}	
\usepackage{amssymb}	
\usepackage{bm}		
\usepackage{url}
\usepackage{amsbsy}
\usepackage{xspace}
\usepackage{hyperref}
\usepackage{longtable}
\usepackage[colorinlistoftodos]{todonotes}
\usepackage[flushleft]{threeparttable}

\usepackage[T1]{fontenc}
\usepackage{ae,aecompl}
\usepackage{chngcntr}

\def\lesssim{\mathrel{\hbox{\rlap{\hbox{\lower5pt\hbox{$\sim$}}}\hbox{$<$}}}}
\def\gtrsim{\mathrel{\hbox{\rlap{\hbox{\lower5pt\hbox{$\sim$}}}\hbox{$>$}}}}

\usepackage{upgreek}                                                  
\usepackage{xspace}                                                       
\newcommand{\um}{$\upmu$m\xspace}            
\newcommand{\ujy}{$\upmu$Jy\xspace} 
\def\apjl{ApJ}%
\def\apjs{ApJS}%
\def\aap{A\&A}%
\shortauthors{Karambelkar et al.}

\begin{document}
\title{Hot springs and dust reservoirs: JWST reveals the dusty, molecular aftermath of extragalactic stellar mergers}

\author[0000-0003-2758-159X]{Viraj Karambelkar}
\email{viraj@astro.caltech.edu}
\affiliation{Cahill Center for Astrophysics, California Institute of Technology, Pasadena, CA 91125, USA}

\author[0000-0002-5619-4938]{Mansi M. Kasliwal}
\affiliation{Cahill Center for Astrophysics, California Institute of Technology, Pasadena, CA 91125, USA}

\author{Ryan M. Lau}
\affiliation{NSF NOIRLab, 950 N. Cherry Ave., Tucson, AZ 85719, USA}
\affiliation{IPAC, Mail Code 100-22, Caltech, 1200 E. California Blvd., Pasadena, CA 91125, USA}

\author[0000-0001-5754-4007]{Jacob E. Jencson}
\affiliation{IPAC, Mail Code 100-22, Caltech, 1200 E. California Blvd., Pasadena, CA 91125, USA}

\author[0000-0003-0901-1606]{Nadejda Blagorodnova}
\affiliation{Institut de Ciències del Cosmos (ICCUB), Universitat de Barcelona (UB), c. Martí i Franquès, 1, 08028 Barcelona, Spain}

\author[0000-0002-3938-4211]{Marco~A. G{\'o}mez-Mu{\~n}oz}
\affiliation{Institut de Ciències del Cosmos (ICCUB), Universitat de Barcelona (UB), c. Martí i Franquès, 1, 08028 Barcelona, Spain}

\author[0000-0001-6314-8131]{Hugo Tranin}
\affiliation{Institut de Ciències del Cosmos (ICCUB), Universitat de Barcelona (UB), c. Martí i Franquès, 1, 08028 Barcelona, Spain}

\author{Maxime Wavasseur}
\affiliation{Institut de Ciències del Cosmos (ICCUB), Universitat de Barcelona (UB), c. Martí i Franquès, 1, 08028 Barcelona, Spain}

\author[0000-0002-9301-5302]{Melissa Shahbandeh}
\affiliation{Department of Physics and Astronomy, Johns Hopkins University, Baltimore, MD 21218, USA}
\affiliation{Space Telescope Science Institute, 3700 San Martin Drive, Baltimore, MD 21218, USA}

\author[0000-0002-8989-0542]{Kishalay De}
\affiliation{Columbia University, 538 West 120th Street 704, MC 5255, New York, NY 10027}
\affiliation{Center for Computational Astrophysics, Flatiron Research Institute, 162, 5th Ave, New York, NY 10010}

\begin{abstract}
    We present \emph{James Webb Space Telescope (JWST)} observations of four Luminous Red Novae (LRNe): dusty, extragalactic transients from stellar mergers following common-envelope evolution (CEE) in massive binary stars. Our targets --- AT\,2021blu, AT\,2021biy, AT\,2018bwo, and M31-LRN-2015 --- span a broad range in progenitor primary masses ($\approx$3--24\,M$_{\odot}$) and post-merger ages ($\approx$1100--3700\,days). All four were observed with the Mid-Infrared Instrument (MIRI) from 5--25\,\um; AT\,2021blu and AT\,2021biy additionally have 5--12\,\um spectra from the Low-Resolution Spectrometer. These spectra show strong features of oxygen-rich molecules, including water vapor, supporting the recent association of water fountain sources with CEE. Radiative transfer modeling of the spectral energy distributions yields dust masses of $\approx$\,4.2$\times10^{-5}$, 3$\times10^{-4}$, 7.5$\times10^{-5}$, and 7.7$\times10^{-4}$\,M$_{\odot}$ respectively --- corresponding to $\approx10$\%, 60\%, 6\% and 12\% of median dust masses in core-collapse supernovae (CCSNe) at similar phases. Accounting for their occurrence rates, we estimate that LRNe can contribute $\sim$25\% as much dust as CCSNe to the cosmic dust budget. Furthermore, the lower expansion velocities of LRNe may reduce dust destruction by reverse shocks compared to CCSNe, potentially increasing this contribution. In addition to dust masses, we use our \emph{JWST} observations to measure late-time properties such as the luminosities, temperatures, radii, and dust-to-gas ratios of the merger remnants. Our results highlight the need for broader infrared studies of LRNe to quantify their contribution to the cosmic dust budget, study the evolution of oxygen-rich molecules, and probe the final fates of CEE. \\
\end{abstract}

\section{Introduction}
Luminous Red Novae (LRNe) are transient eruptions associated with stellar mergers as a consequence of common-envelope evolution (CEE) in binary stars \citep{Kulkarni07, Kasliwal2011, Ivanova13, Pastorello2019a,Blagorodnova2021}. These transients provide a valuable opportunity to study CEE --- a crucial but poorly understood phase in binary evolution that plays a key role in the formation of double compact objects, which are important sources of gravitational waves \citep{Ivanova2013araa, VignaGomez2020, Dominik12, Postnov2014, Marchant2021}. Additionally, LRNe are prolific dust producers, and offer insights into the contribution of CEE events to the cosmic dust budget \citep{Lu2013, Iaconi2020, Bermudez2024}.

LRNe span a wide range of peak luminosities ($-3 \geq $\,M$_{r} \geq -15$), which are correlated with the mass of the primary star \citep{Kochanek14_mergers, Pastorello2019a, Blagorodnova2021}. The lower-luminosity members of this class, often referred to as red novae (RNe), are primarily Galactic events with primary star masses between $\approx 1-8$\,M$_{\odot}$ \citep{Tylenda11, Munari2002AA, Tylenda05b, Tylenda13}.  In contrast, more luminous LRNe (M$_{\rm{r, peak}}\leq-10$) are extragalactic transients involving primary stars more massive than 10\,M$_{\odot}$ \citep{Blagorodnova2021}. Despite their broad luminosity range, these transients exhibit similar late-time behaviors, marked by a shift of the emitted light to infrared (IR) wavelengths due to dust-formation and spectra resembling M-type supergiant stars with strong absorption features due to oxygen-rich molecules \citep{Blagorodnova2020, Karambelkar2023, Tylenda05a, Banerjee2004}. Furthermore, LRNe have high occurrence rates, with Galactic low mass red novae having an estimated rate of $\sim0.5$\,yr$^{-1}$ \citep{Kochanek14_mergers} and extragalactic massive LRNe occurring at a volumetric rate that is $\approx77$\% of the local core-collapse supernova (CCSN) rate  \citep{Karambelkar2023}. 

Given their high occurrence rates and dusty nature, LRNe may be significant contributors to the cosmic dust budget \citep{Lu2013, Bermudez2024}. Late-time IR and sub-mm observations of the Galactic RNe show that their remnants are enshrouded in substantial amounts of dust and molecules \citep{Nicholls2013, Lynch2007, Banerjee2004, Banerjee2007, Woodward2021, Steinmetz2025}. The more luminous extragalactic LRNe have larger ejected masses (e.g., \citealt{Matsumoto2022}) and are thus expected to produce even more dust than their Galactic counterparts. Given their volumetric rate is comparable to CCSNe -- considered to be major cosmic dust sources \citep{Nozawa2003, Sarangi2015} -- these LRNe could represent a similarly important dust production channel. Moreover, while CCSN-generated dust can be partially destroyed by reverse shocks propagating through their ejecta, the lower velocities of LRNe suggest a reduced destruction fraction (e.g. \citealt{Nozawa2007, Slavin15, Slavin2020}). Thus, extragalactic LRNe could be major contributors to the cosmic dust budget and could explain unusually dusty galaxies in the early universe \citep{Dwek2007}. However, while a handful of mid-IR observations exist for Galactic LRNe, the emerging population of extragalactic LRNe has not been characterized at mid-IR ($>10$\,\um) wavelengths. 

In addition to dust, late-time observations of LRNe can shed light on the formation and evolution of molecules as they condense to form ice and dust. IR and sub-mm observations of Galactic RNe years after their eruptions reveal complex, bipolar ejecta structures rich in gaseous molecules such as water vapor, CO, SiO, AlO  \citep{Kaminski2011AA, Kaminski2021, Kaminski2018, Steinmetz2024} as well as conglomerates such as water ice \citep{Banerjee2004}. However the molecular aftermath of the more massive LRNe has not been studied. Late-time observations also provide crucial insights about the CEE phase, as the dust masses can be used to infer whether CEE resulted in a stellar merger or an envelope ejection \citep{MacLeod17, Blagorodnova17}. Additionally, late-time rebrightenings in LRNe \citep{Cai2022} are believed to be powered by interactions of the ejecta with circumstellar material (CSM), and provide a way to trace pre-merger mass loss during CEE. Despite their importance, late-time observations of the extragalactic LRNe have been limited in the past due to a lack of sensitive IR telescopes.

In this paper, we present late-time mid-IR observations of four LRNe obtained with the \emph{James Webb Space Telescope}, including the first mid-IR spectroscopic observations of extragalactic LRNe. We model the observations to measure the dust masses produced in these events and estimate their contribution to the cosmic dust budget relative to CCSNe. We also study the molecular content of the remnants and use the \emph{JWST} observations to probe the late-time evolution of massive stellar merger remnants. We describe the properties of the four LRNe in Section \ref{sec:lrn_description}, our \emph{JWST} observations in Section \ref{sec:observations}, our modeling in Section \ref{sec:analysis}, and discuss the implications of our findings in Section \ref{sec:discussion}. We conclude with a summary of our results in Section \ref{sec:summary}.

\section{The Luminous Red Novae}
\label{sec:lrn_description}
Four LRNe are presented in this paper -- AT\,2021blu \citep{Pastorello2022, Karambelkar2023}, AT\,2021biy \citep{Cai2022, Karambelkar2023}, AT\,2018bwo \citep{Blagorodnova2021}, and M31-LRN-2015 \citep{Williams2015, MacLeod17}. 

AT\,2021blu and AT\,2021biy were both discovered in early 2021 in the galaxies UGC\,5829 (9.5 Mpc \footnote{All distances taken from ned.ipac.caltech.edu, corrected for infall into the Virgo, GA, and Shapley clusters and assuming H$_{0}=73$\,km\,s$^{-1}$\,Mpc$^{-1}$}) and NGC\,4631 (7.7 Mpc) respectively. Their peak absolute magnitudes were M$_{\rm{r, peak}} = -13.50 \pm 0.15$ and $-13.86 \pm 0.15$ respectively, putting them at the luminous end of the LRN-luminosity function.  Their lightcurves showed similar initial blue peaks lasting $\approx$50 days followed by a long plateau lasting a few hundred days. While this plateau in AT\,2021blu lasted $\approx 180$ days, the plateau-duration in AT\,2021biy exceeded 250 days, substantially longer than seen in other LRNe with similar luminosities. The maximum light spectra show strong and narrow ($v \leq 1000$\,km\,s$^{-1}$) emission lines of hydrogen. At late times, both transients evolve show strong infrared excesses \citep{Karambelkar2023} indicating dust-formation, and their spectra show the characteristic broad absorption bands of oxygen-rich molecules such as water vapor, TiO, CO, and VO . Archival pre-eruption imaging shows that the progenitor systems of both these transients involved luminous yellow supergiant stars as donor stars, with masses between $\approx 13-18$\,M$_{\odot}$ for AT\,2021blu \citep{Pastorello2022} and $\approx17-24$\,M$_{\odot}$ for AT\,2021biy \citep{Cai2022}. 

AT\,2018bwo and M31-LRN-2015 were discovered in May 2018 and January 2015 in the galaxies NGC\,45 (6.7\,Mpc) and M\,31 (0.77\,Mpc) respectively. The transients reached peak absolute magnitudes of M$_{\rm{r, peak}} = -10.97 \pm 0.11$ and $-9.5\pm0.1$ respectively -- placing them at the low luminosity end of extragalactic LRNe. In addition to their lower luminosities compared to AT\,2021blu and AT\,2021biy, they also had shorter durations, with initial blue peaks lasting $\sim10$ days and subsequent plateaus lasting $\approx$ 41 days each. Despite this, they show characteristic LRN signatures with low expansion velocities ($\sim$100\,km\,s$^{-1}$) and late-time spectra that resemble M-type stars with molecular absorption features. Archival imaging revealed the progenitor primary masses of 12--16\,M$_{\odot}$ for AT\,2018bwo, and 3--5\,M$_{\odot}$ for M31-LRN-2015. 

\begin{figure}
    \centering
    \includegraphics[width=0.45\textwidth]{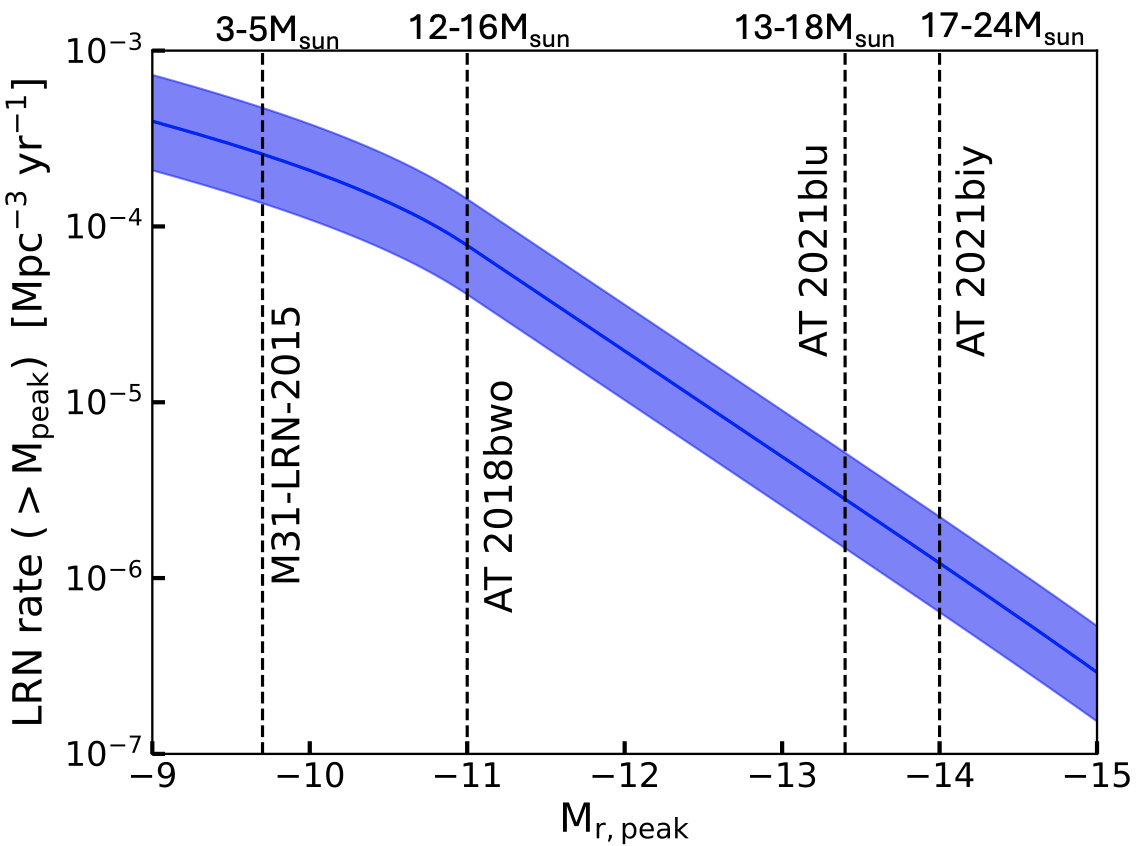}
    \caption{The cumulative volumetric rate of LRNe as a function of their peak absolute magnitudes, adapted from \citet{Kochanek14_mergers,Karambelkar2023}. The four LRNe in our sample are indicated by vertical lines. Ranges of primary progenitor masses are indicated on the top.}
    \label{fig:lrn_sample}
\end{figure}

The differences in luminosities and durations observed in these LRNe have been attributed to different ejecta masses \citep{Matsumoto2022} which likely correlate with the progenitor masses \citep{Blagorodnova2021}. The broad range of occurrence rates, luminosities, progenitor masses, and phases since eruption spanned by these four LRNe (see Figure \ref{fig:lrn_sample}) allow for studying dust-formation across a wide range of the stellar merger phase space.

\section{Observations and Data reduction}
\label{sec:observations}
All \emph{JWST} observations of the four LRNe were conducted as part of the GO Program 4244 (PI Karambelkar).
\subsection{AT\,2021blu}
\subsubsection{JWST observations}
\label{sec:2021blu_jwst_observations}
AT\,2021blu was observed with the Mid-Infrared Instrument (MIRI, \citealp{Wright2023}) aboard the \emph{JWST} on 2024 April 3. The observations comprised of a R$\sim100$ spectrum covering 5-12 \um obtained with the Low-Resolution Spectrometer (LRS) and broadband imaging observations in the F1280W, F1500W, F1800W, F2100W and F2550W filters. Calibrated data were downloaded from the MAST portal. The LRS data were reduced using version 1.13.3 of the JWST science calibration pipeline and calibrated using version 11.17.14 of the CRDS with the 1223 CRDS context. The LRS spectra were extracted using the optimal extraction algorithm \citep{Horne1986} and LRS observations of the flux standard BD+60~1753 as a reference profile. The optimal extraction was implemented by adapting routines in the JWST data analysis tool notebook for MIRI LRS observations\footnote{\url{https://spacetelescope.github.io/jdat_notebooks/}}. 

For the final spectral extraction of AT\,2021blu, we used an extraction aperture with a full width of 11 pixels and background regions between 6 and 11 pixels from the center of the profile trace on either side. The background at the central trace position was determined using the average of the fluxes in the background regions. AT\,2021blu is offset from its host galaxy and is relatively isolated, so we do not detect any structure in the background emission in the 2D-spectrum.  

For the broadband imaging, we performed aperture photometry at the locations of AT\,2021blu (RA=160.643049, Dec = +34.437396) on the MIRI images using an aperture that encloses 70\% of the total flux and aperture corrections from \texttt{jwst\_miri\_apcorr\_0010.fits}\footnote{Downloaded from \href{https://jwst-crds.stsci.edu/}{https://jwst-crds.stsci.edu/}}. 

\subsubsection{NEOWISE mid-IR observations}
The field of AT\,2021blu was observed multiple times by NEOWISE \citep{Mainzer2014ApJ} in the W1 (3.6\,\um) and W2 (4.5\,\um) bands as part of its survey on 2024 April 27 (24 days after the \emph{JWST} observations). We stacked the NEOWISE frames from using the online NEOWISE image coadding service\,\footnote{https://irsa.ipac.caltech.edu/applications/ICORE/}. We constructed W1 and W2 reference images by coadding observations of the field from 2018. We subtracted the new images from the reference images using the ZOGY algorithm \citep{Zackay2016} and performed PSF photometry at the location of AT\,2021blu in the difference image. The measured flux values are listed in Table \ref{tab:at2021blu_fluxes}. 

\begin{figure*}
    \centering
    \includegraphics[width=0.24\textwidth]{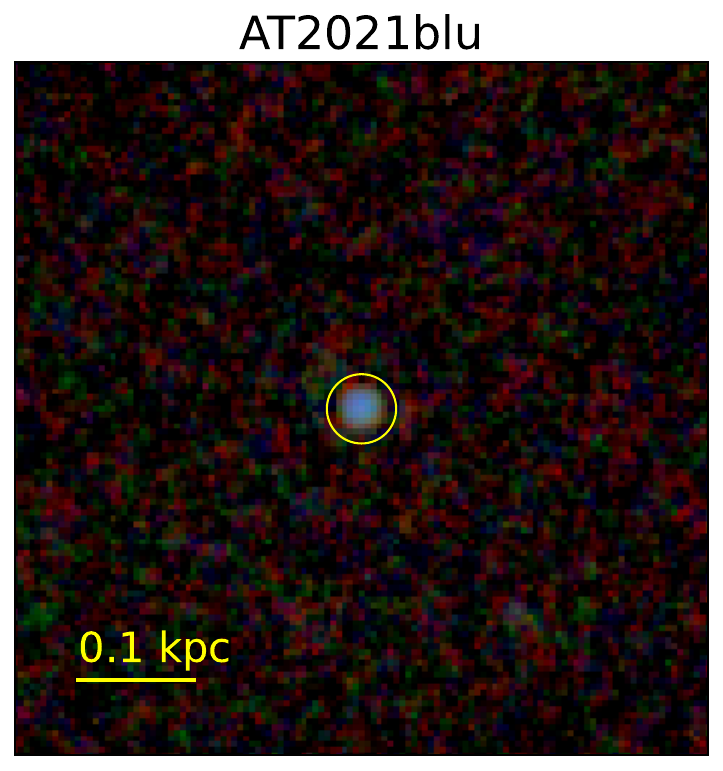}\includegraphics[width=0.24\textwidth]{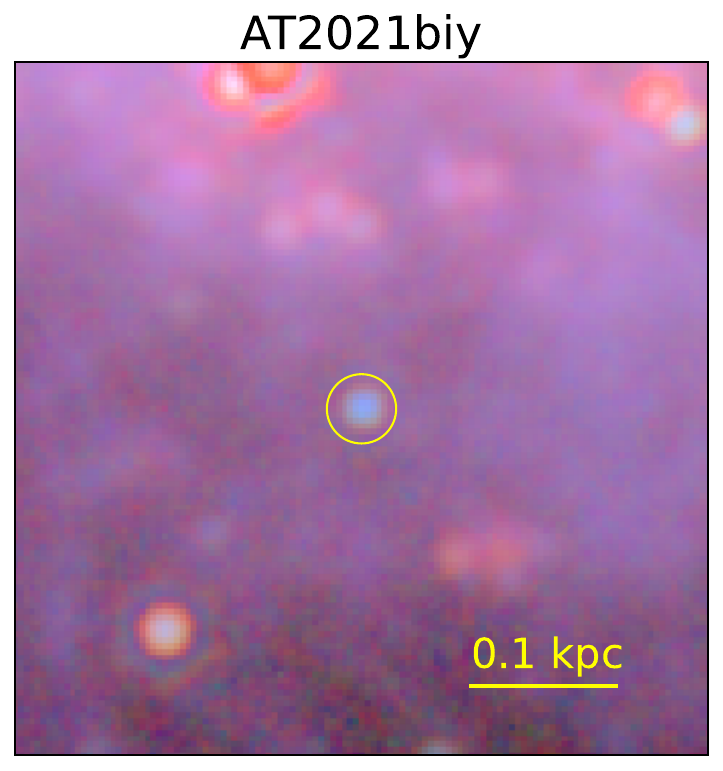}
    \includegraphics[width=0.24\textwidth]{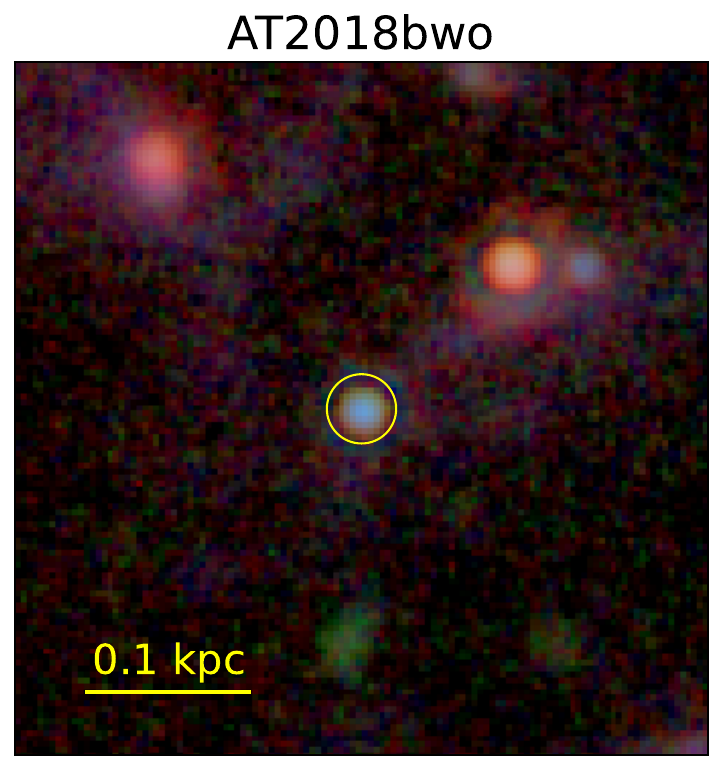}\includegraphics[width=0.24\textwidth]{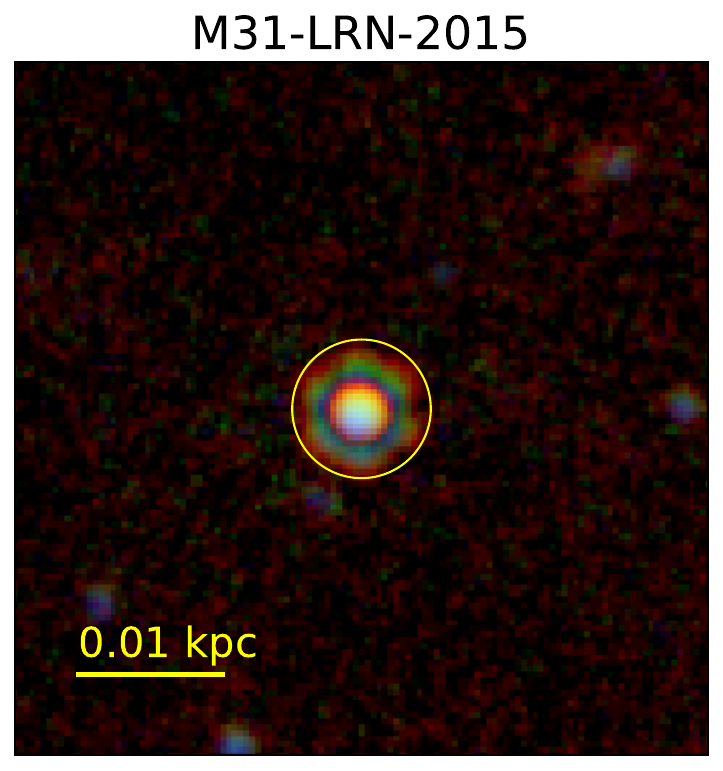}\\
    \caption{\emph{JWST} MIRI three-color images of the four LRNe presented in this paper. The color scheme used is blue: F1280W, green: F1500W, and red: F1800W. The yellow circles mark the positions of the LRNe.}
    \label{fig:jwst_color_image}
\end{figure*}
\subsubsection{NIR observations}
We observed AT\,2021blu in the NIR \emph{J}, \emph{H} and \emph{Ks} bands with the Wide-field Infrared Camera (WIRC, \citealt{Wilson2003}) on the 200-inch Hale telescope on 2024 June 5. The WIRC data were reduced using standard methods for dark subtraction, flat-fielding, sky subtraction, astrometric and photometric calibration \citep{De2021} and PSF photometry was performed at the location of AT\,2021blu to extract the fluxes. The NIR flux measurements are listed in Table \ref{tab:at2021blu_fluxes}. 

We also obtained a NIR 1-2.5\,\um spectrum of AT\,2021blu with the Near-infrared Echelle Spectrograph (NIRES, \citealt{Wilson2004}) on the Keck II 10\,m telescope on 2024 June 16. The spectra were reduced, extracted and telluric corrected using the software \texttt{spextool} \citep{Vacca2003, Cushing2004}. 

\begingroup
\renewcommand{\tabcolsep}{10pt}
\begin{table}
\begin{center}
\begin{minipage}{8cm}
\caption{Infrared fluxes of the four LRNe}
\label{tab:at2021blu_fluxes}
\begin{tabular}{lccccc}
\hline
\hline
{Filter} & {Date} & {Flux} \\
{}       &   {}   & {\ujy}    \\
\hline
 & AT\,2021blu & \\
\hline
J      & 2024-06-05 & 7.0 $\pm$ 1.8 \\
H      & 2024-06-05 & 15.4 $\pm$ 3.1\\
Ks     & 2024-06-05 & 42.0 $\pm$ 4.2\\
W1     & 2024-04-27 & 150 $\pm$ 15\\
W2     & 2024-04-27 & 93 $\pm$ 23 \\
F560W  & 2024-04-03 & 116.4 $\pm$ 1.2\\
F1280W & 2024-04-03 & 100.7 $\pm$ 2.2\\
F1500W & 2024-04-03 & 89.3 $\pm$ 2.7\\
F1800W & 2024-04-03 & 73.1 $\pm$ 5.0\\
F2100W & 2024-04-03 & 59.7 $\pm$ 6.8\\
F2550W & 2024-04-03 & 51.0 $\pm$ 17.1\\
\hline
 & AT\,2021biy & \\
\hline
 F560W  & 2024-06-08 & 252.9 $\pm$ 7.0\\
F1280W & 2024-06-08 & 359.0 $\pm$ 9.7\\
F1500W & 2024-06-08 & 309.2 $\pm$ 8.4\\
F1800W & 2024-06-08 & 293.9 $\pm$ 10.9\\
F2100W & 2024-06-08 & 274.4 $\pm$ 13.0\\
F2550W & 2024-06-08 & 247.0 $\pm$ 24.6\\
\hline 
 & AT\,2018bwo & \\
\hline
F560W & 2024-07-04 & 86.0 $\pm$ 1.0\\
F770W & 2024-07-04 & 114.1 $\pm$ 1.2\\
F1000W & 2024-07-04 & 159.8 $\pm$ 1.3\\
F1130W & 2024-07-04 & 172.9 $\pm$ 3.5\\
F1280W & 2024-07-04 & 170.3 $\pm$ 2.1\\
F1500W & 2024-07-04 & 167.9 $\pm$ 2.4\\
F1800W & 2024-07-04 & 154.9 $\pm$ 3.6\\
F2100W & 2024-07-04 & 134.6 $\pm$ 4.3\\
F2550W & 2024-07-04 & 95.6 $\pm$ 16.5\\
\hline 
& M31-LRN-2015 & \\
\hline
F560W & 2025-01-16 & 46.2 $\pm$ 0.9\\
F770W & 2025-01-16 & 227.0 $\pm$ 0.9\\
F1000W & 2025-01-16 & 51.3 $\pm$ 1.2\\
F1130W & 2025-01-16 & 162.4 $\pm$ 3.2\\
F1280W & 2025-01-16 & 551.1 $\pm$ 2.7\\
F1500W & 2025-01-16 & 740.8 $\pm$ 3.8\\
F1800W & 2025-01-16 & 755.8 $\pm$ 4.6\\
F2100W & 2025-01-16 & 1069.2 $\pm$ 5.3\\
F2550W & 2025-01-16 & 1490.4 $\pm$ 17.9\\
\hline

\end{tabular}

\end{minipage}
\end{center}
\end{table}
\endgroup

\begin{figure*}[hbt]
    \centering
    \includegraphics[width=0.49\textwidth]{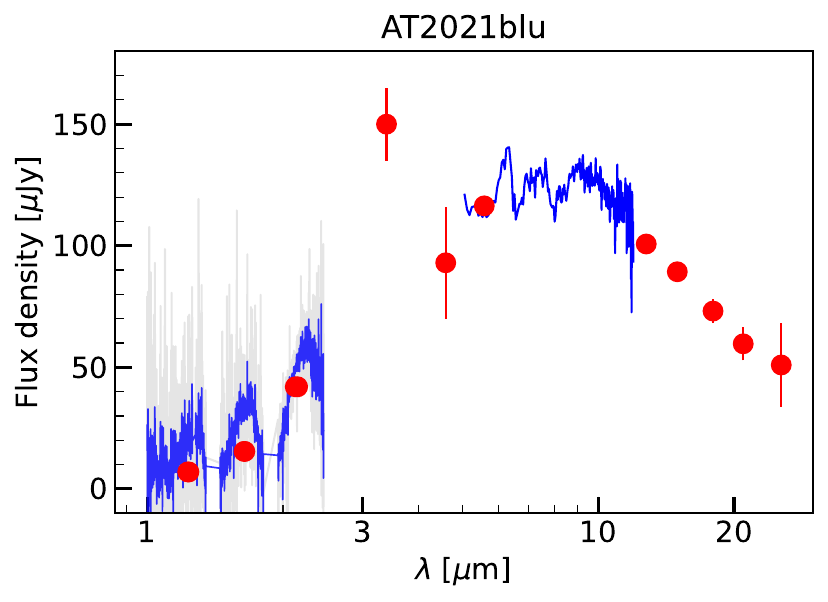}
    \includegraphics[width=0.49\textwidth]{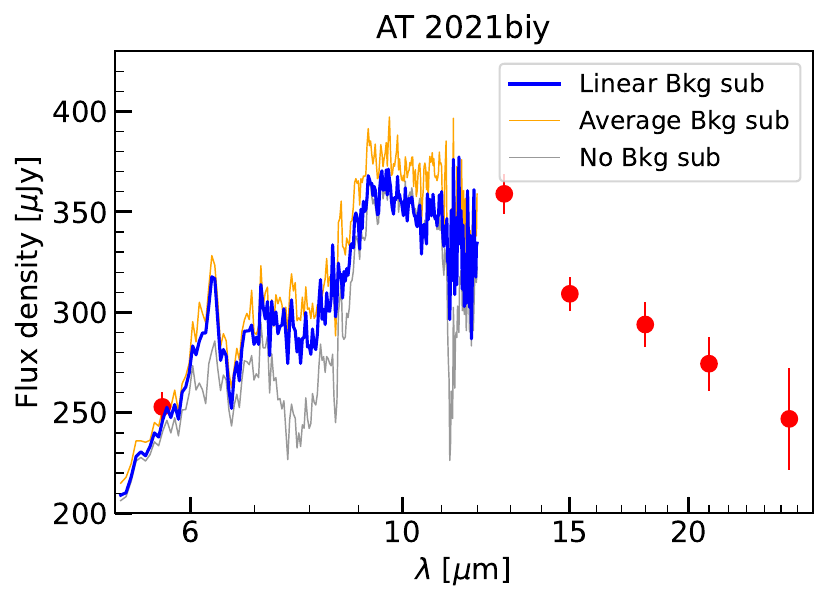}\\ \includegraphics[width=0.5\textwidth]{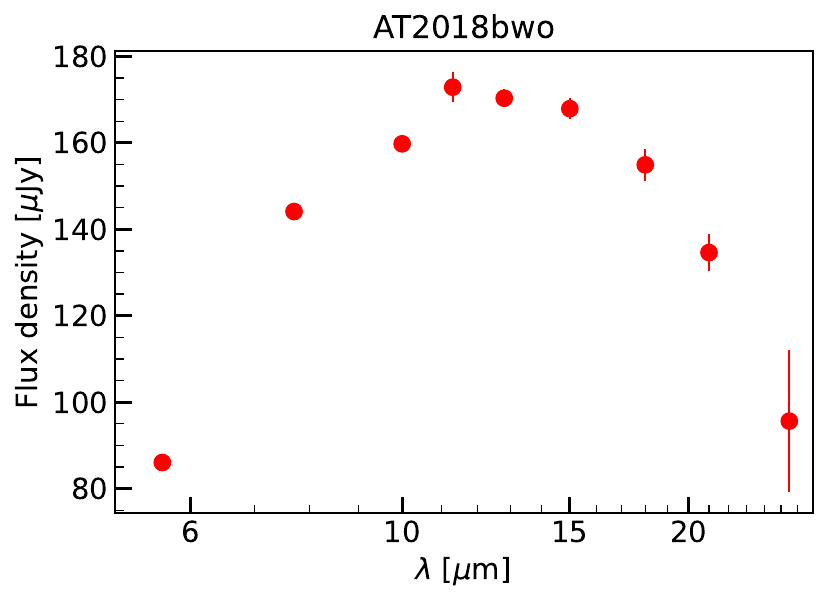}\includegraphics[width=0.5\textwidth]{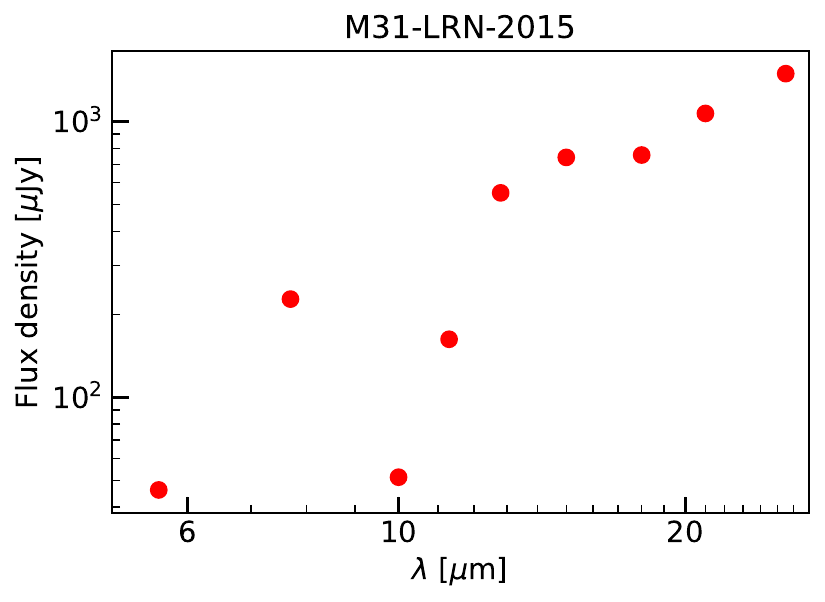}
    \caption{The infrared spectral energy distributions of the four LRNe. For AT\,2021biy \emph(top right), three reductions of the LRS spectrum are plotted with different methods of background subtraction (see text). In the rest of the paper, the reduction with linear background subtraction (blue) is adopted.
 }
    \label{fig:lrne_ir_seds}
\end{figure*}

\subsection{AT\,2021biy}
\emph{JWST} observations of AT\,2021biy were conducted on 2024 June 8 and comprised of a 5-12\um MIRI LRS spectrum and MIRI imaging in the F1280W, F1500W, F1800W, F2100W, and F2550W filters. The spectra and images were reduced using the same methods described in Section \ref{sec:2021blu_jwst_observations}. 

In contrast to AT\,2021blu, AT\,2021biy is located on top of a star-forming region within its host galaxy. Consequently, the 2D-spectrum shows a complex background profile, especially in the 8--10\um range. For AT\,2021biy, we used background regions between 5 and 13 pixels from the center of the profile trace on either side. We determined the background at the central trace location by fitting a linear function to the flux profiles in the left and right background regions. Figure \ref{fig:lrne_ir_seds} shows a comparison of the extracted spectra of AT\,2021biy with this linear-background subtraction, an average-background subtraction (similar to AT\,2021blu) and without any background subtraction. Without background subtraction, the flux from AT\,2021biy is significantly underestimated between 8--10\,\um. The negative background in this region likely results from over-subtraction during the 2D background subtraction performed by the pipeline during the reduction stage. The linear and average-background subtracted spectra are broadly consistent with each other and only show slight differences in the profiles of the features. We use the linear-background subtracted spectrum for our analysis in this paper, but note that it is possible that some flux measurements in this wavelength range could be affected by the uncertain background levels.

For the imaging observations, we performed aperture photometry at the location of AT\,2021biy (RA = 190.516707, Dec = +32.535532) using the process described in Section \ref{sec:2021blu_jwst_observations}.

\subsection{AT\,2018bwo and M31-LRN-2015}
The MIRI observations of AT\,2018bwo and M31-LRN-2015 were conducted on 2024 July 04, and 2025 January 16 respectively, and comprised of broadband imaging with the F560W, F770W, F1000W, F1130W, F1280W, F1500W, F1800W, F2100W, and F2550W filters for both sources. The data were processed and fluxes were measured using the same method described in Section \ref{sec:2021blu_jwst_observations} by performing aperture photometry at the locations of AT\,2018bwo (RA=3.507175, Dec=$-23.1932145$) and M31-LRN-2015 (RA=10.533503, Dec=40.9169811). 

The measured broadband fluxes and uncertainties are listed in Table \ref{tab:at2021blu_fluxes}. 

\section{Analysis}
\label{sec:analysis}
\subsection{SEDs and Comparison with Galactic mergers}
Figure \ref{fig:lrne_ir_seds} shows the IR spectral energy distributions (SEDs) of the four LRNe. AT\,2021blu and AT\,2021biy have spectroscopic and photometric observations. The 5--12\,\um spectra of both these sources are dominated by strong features at 6.3\,\um, 7.5\,\um and a broad feature spanning 8--10\,\um. In the following sections, we show that these features can be explained by a combination of O-rich dust and molecules around the merger remnant. The NIR spectrum of AT\,2021blu also shows strong absorption features due to TiO and water vapor. The AT\,2021blu SED peaks around 5\,\um, while AT\,2021biy peaks around 10\,\um, suggesting a colder dust temperature. AT\,2018bwo and M31-LRN-2015 have only photometric observations, so we cannot identify any circumstellar molecules in them. The SED of AT\,2018bwo peaks around 11\um and declines at longer wavelengths, similar to AT\,2021biy. In contrast, the SED of M31-LRN-2015 rises sharply from 15 to 25 \um, indicating much colder dust. The SED also shows a deep absorption feature at 10\,um, generally attributed to silicate dust.

\begin{figure*}
    \centering
    \includegraphics[width=0.8\textwidth]{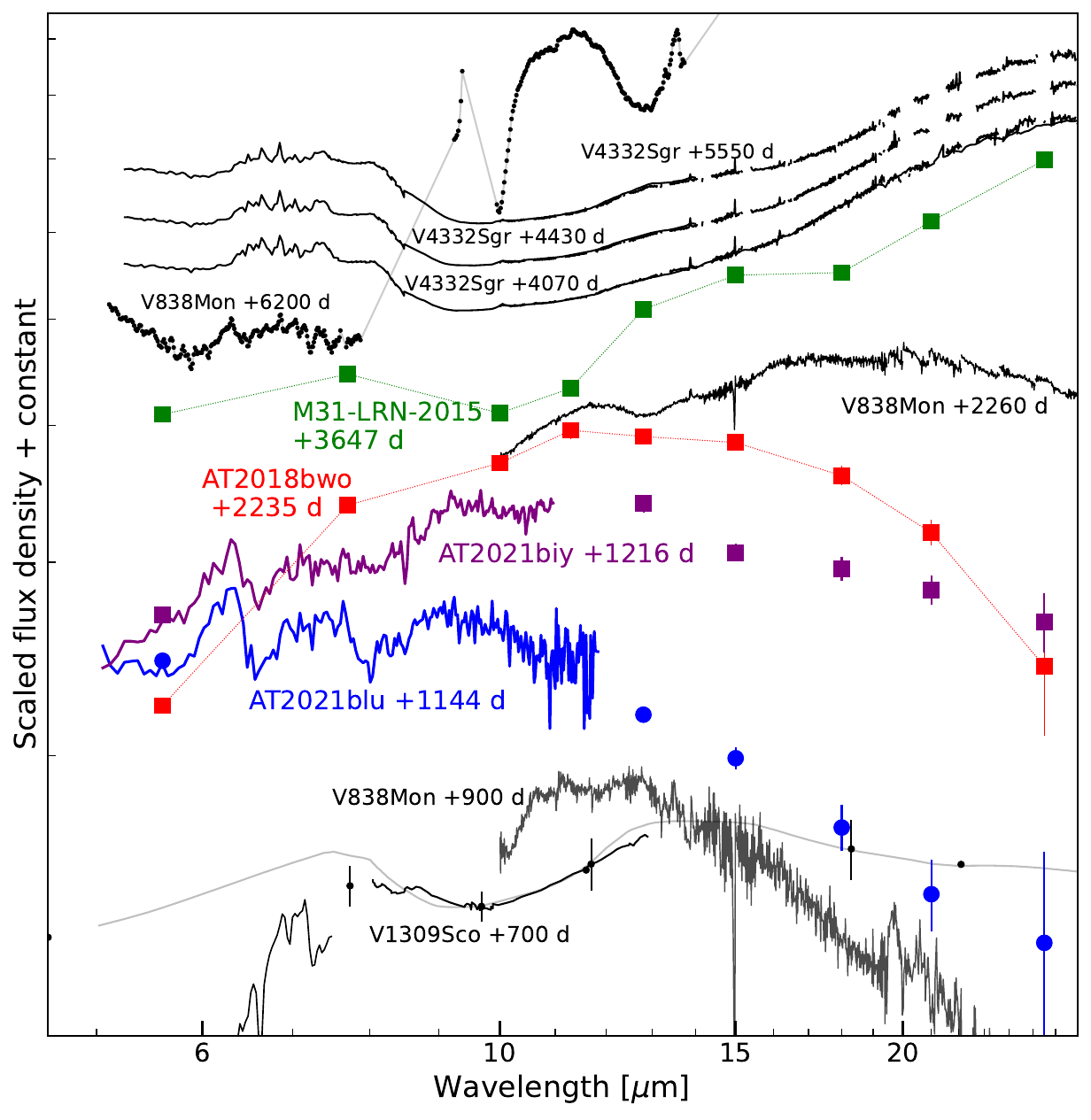}
    \caption{Comparison of the spectral energy distributions of the four LRNe (colored) to Galactic stellar mergers (black). }
    \label{fig:at2021blu_comparisons}
\end{figure*}

Figure \ref{fig:at2021blu_comparisons} compares the \emph{JWST} mid-IR SEDs of the four LRNe to the Galactic stellar mergers V1309Sco (taken from \citealp{Nicholls2013}), V4332Sgr \citep{Banerjee2007} and V838Mon \citep{Woodward2021}. In addition to these published data, we also downloaded reduced archival Spitzer Infrared Spectrograph (IRS; \citealp{Houck2004}) spectra for V838Mon and V4332Sgr from the Combined Atlas of Sources with Spitzer IRS Spectra (CASSIS\,\footnote{\href{https://cassis.sirtf.com}{https://cassis.sirtf.com}; AORKeys : 10523136, 25433344, 10523392, 14867968, 25432064}; \citealp{Lebouteiller2011, Lebouteiller2015}). AT\,2021blu and AT\,2021biy were observed with \emph{JWST} at 1144 and 1216 days since their optical \emph{r-}band peaks. The shape of the mid-IR SEDs of AT\,2021blu and AT\,2021biy bear closest resemblance to that of V838Mon at a phase of 900 days since peak. The 6.3\,\um feature of AT\,2021blu and AT\,2021biy is also seen in the 6200 day spectrum of V838\,Mon and has been attributed to water vapor \citep{Woodward2021},  similar to T Tauri Stars \citep{Sargent2014}. Notably absent from their spectra are the strong, broad silicate absorption from 8--10\,\um that dominate the mid-IR spectra of V1309Sco at a phase of 700 days since peak, and V838\,Mon and V4332\,Sgr at phases $>2000$\,days.  AT\,2018bwo and M31-LRN-2015 were observed at later phases of 2235 days and 3647 days since their respective peaks. AT\,2018bwo shows a sharp rise from 5 to 8\,\um that is steeper than the other three LRNe, indicative of an emission feature. In the next section, we find that this behavior can be explained by silicate emission. M31-LRN-2015 shows a strong silicate absorption feature spanning 7 to 13\,\um similar to the 700 day SED of V1309\,Sco and the very late time SEDs of V4332\,Sgr and V838\,Mon. Similar to these events, the SED of M31-LRN-2015 rises sharply towards wavelengths longer than 25\,\um.

\subsection{SED modeling}
\label{sec:model_fitting}
We now model the infrared SEDs of the four LRNe to constrain their dusty and molecular content. A detailed 3D radiative transfer analysis is outside the scope of this paper. Here, we construct spherically symmetric models comprising a central star surrounded by shells of dust and molecules, using the radiative transfer code \texttt{DUSTY} \citep{Ivezic97,Ivezic99}. For the two sources with spectroscopic observations, we include an additional spherical shell with molecules to constrain column densities of different molecules.
\subsubsection{AT\,2021blu and AT\,2021biy}
\label{sec:blu_biy_model_fitting}
As the infrared spectra of AT\,2021blu and AT\,2021biy are dominated by strong molecular features, we attempt to model the SED as a combination of a central star surrounded by a dust shell and a second outer molecular shell. This geometry is motivated by ALMA observations of the Galactic stellar mergers \citep{Kaminski2018, Kaminski2021}, which show that the remnants are enshrouded in dust and molecules, with the molecules generally extending to larger radii than the dust. Under this assumption, we first model radiation from the star emerging through a dust shell using the radiative transfer code \texttt{DUSTY}. Then we model the emergent radiation when the \texttt{DUSTY} output passes through the outer molecular shell to derive the column densities of molecules encountered in the outer regions of the CSM. A full description and calculation of our model is given in Appendix A.

For the \texttt{DUSTY} models, following \citet{Blagorodnova2020}, we assumed a standard \citet{Mathis1977} distribution for dust grain sizes that assumes minimum and maximum grain sizes of 0.005 $\mu$m and 0.25 $\mu$m, respectively, and a grain-size power-law index of $-3.5$. We assumed a r$^{-2}$ radial density profile for the dust shell and a fixed shell-thickness ratio of 2. For the dust composition, we assumed a mixture of silicate and compact alumina dust using the \texttt{Sil-DL} and \texttt{Al2O3-comp} opacities available in \texttt{DUSTY}. We assumed a silicate-to-alumina ratio of 0.5 (similar to V\,838Mon; \citealp{Woodward2021}). 

For the molecular modeling, we adopt a simple approach commonly used in stellar atmospheric modeling (see e.g. \citealp{Yamamura1999b,Yamamura1999a,Cami2002, Sargent2014}) in which the atmosphere is modeled using the plane-parallel approximation under local-thermodynamic equilibrium (LTE). In this approach, the molecular layer is characterized by four parameters -- the excitation temperature, radius, column density and turbulent velocity. For our modeling, we assume a single circular molecular slab of temperature T$_{\rm{mol}}$ and radius R$_{\rm{mol}}$ comprising CO, H$_{2}$O and SiO molecules with column densities N$_{\rm{CO}}$, N$_{\rm{H}_{2}\rm{O}}$ and N$_{\rm{SiO}}$ respectively placed in front of the star and its dust shell. For AT\,2021biy, we do not include CO as we do not have any NIR data to constrain this molecule. The absorption cross-sections for H$_{2}$O and CO were calculated using data from the HITEMP online database\,\footnote{\href{https://hitran.org/hitemp}{https://hitran.org/hitemp}}  \citep{Rothman2010}, and from the Exomol database for SiO\,\footnote{\href{https://www.exomol.com/data/molecules/SiO/28Si-16O/SiOUVenIR}{https://www.exomol.com/data/molecules/SiO/28Si-16O/SiOUVenIR/}} \citep{Yurchenko2021}. The turbulent velocity of this layer is fixed to v$_{\rm{turb}}=10$\,km\,s$^{-1}$. 

Our choice of these three molecules is motivated by their common occurrence in O-rich environments such as M-type stars and their high opacities in the wavelengths covered by our observations (see \citealp{Sloan2015} for mid-IR spectra of M giants with these molecular features). All three molecules have been identified in the near and mid-infrared spectra of V838\,Mon \citep{Lynch04, Woodward2021}. Furthermore, the NIR spectra of AT\,2021blu shows strong absorption features due to water vapor. Although we do not have NIR observations of AT\,2021biy contemporaneous with the \emph{JWST} observations, an older NIR spectrum presented in \citep{Karambelkar2023} shows similar water-vapor and CO absorption. It is therefore natural to include these molecules in the analysis of the mid-IR spectra. Water vapor has several strong transitions in the 5--8\um range and are expected to significantly absorb the continuum radiation in this region. The strong apparent emission feature peaking at  6.3\,\um in both our LRNe is actually caused by a minimum of water vapor opacity in the narrow wavelength range from 6--7\,\um. In addition to AT\,2021blu and AT\,2021biy, this feature has been detected in V838\,Mon, several RSG stars \citep{Tsuji2002} and outbursting FU-Ori stars \citep{Sargent2014}. SiO gas is also an important ingredient of molecules around RSGs and is responsible for the dip seen at 8\,\um in our spectra. CO gas is responsible for the dip seen between 4-5\,\um in AT\,2021blu, however, we cannot place tight constraints on its column density as the NEOWISE detections have low signal-to-noise ratios. Furthermore, as the SED of AT\,2021biy does not cover these wavelengths, we exclude the CO gas from the model for this source.

\begin{figure*}
    \centering
    \includegraphics[width=0.5\textwidth]{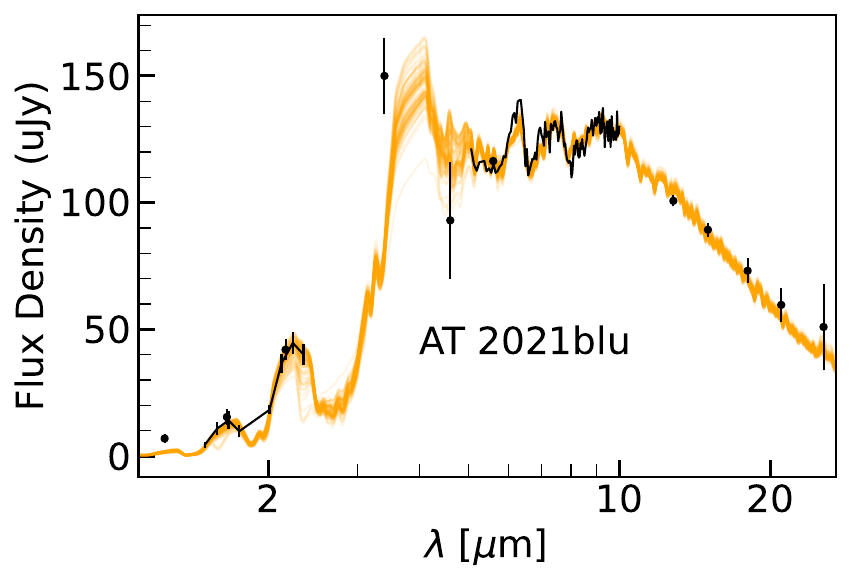}\includegraphics[width=0.5\textwidth]{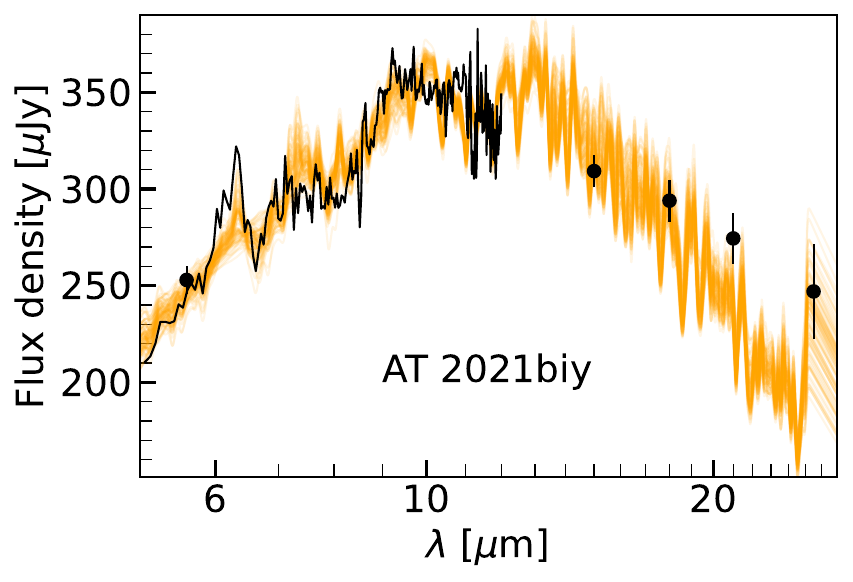}\\
    \includegraphics[width=0.5\textwidth]{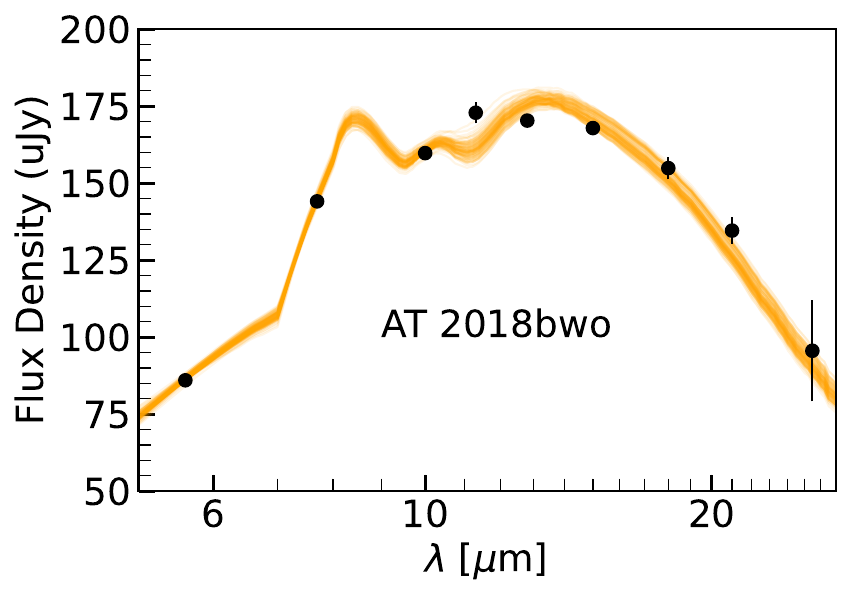}\includegraphics[width=0.5\textwidth]{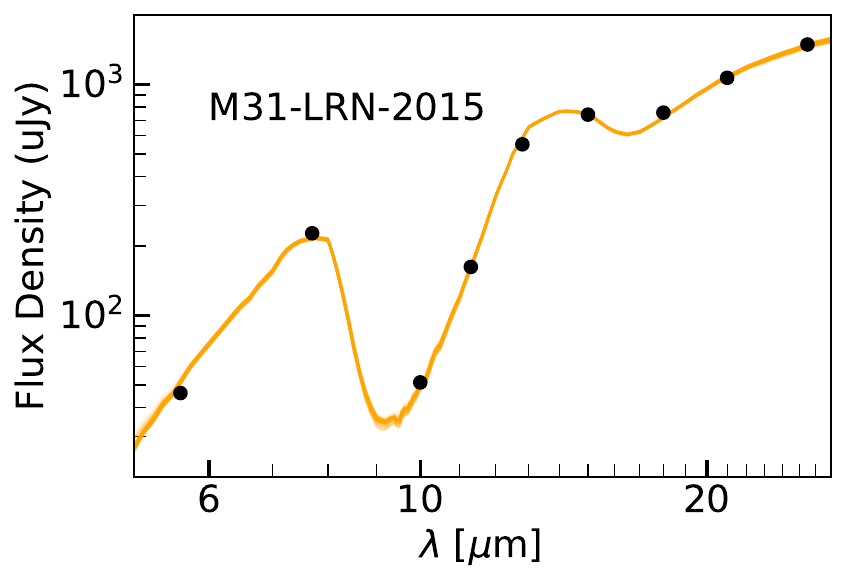}
    \caption{Range of best-fit models (orange) from MCMC fits to the SEDs (black dots). }
    \label{fig:best_fit_models}
\end{figure*}

In total, our model comprises eight free parameters (three \texttt{DUSTY} and five molecular) for AT\,2021blu and seven free parameters (three \texttt{DUSTY} and four molecular) for AT\,2021biy. We use Markov Chain Monte Carlo (MCMC) sampling implemented in the \texttt{python} package \texttt{emcee} \citep{Mackey-Foreman2013} to derive posterior distributions on these parameters, using wide and uniform priors. The resulting medians, 16$^{th}$ and 84$^{th}$ percentile values for the derived parameters are listed in Table \ref{tab:mcmc_results}. Figure \ref{fig:best_fit_models} shows the best-fit models for the two sources. The individual contributions of dust and molecules are shown in Figure \ref{fig:at2021blu_component_fits} and the posterior distributions on all parameters are shown in Figures \ref{fig:corner_at2021blu} and \ref{fig:corner_at2021biy} in Appendix A.   

\begingroup
\renewcommand{\tabcolsep}{10pt}
\begin{table*}
\begin{center}
\begin{minipage}{14cm}
\caption{Inferred dust and molecular parameters for the four LRNe}
\label{tab:mcmc_results}
\begin{tabular}{ccccccccccc}
\hline
\hline
{Source} & AT\,2021blu & AT\,2021biy & AT\,2018bwo & M31-LRN-2015\\
\hline
{Peak MJD}  & 59259 & 59253 & 58260 & 57044 \\
{Phase [d]} & +1144 & +1216 & +2235 & +3647 \\
{Distance (Mpc)} & 9.5 & 7.7 &  6.74  & 0.77      \\
{T$_{\rm{star}}$} [K]  &  2200$_{-200}^{+200}$ & 2400$_{-600}^{+600}$ & 2600$_{-300}^{+400}$ & 6000$_{-300}^{+500}$\\

{T$_{\rm{dust}}$} [K]  & 1100$_{-50}^{+50}$ & 700$_{-70}^{+70}$ & 670$_{-30}^{+30}$ & 385$_{-10}^{+10}$\\

{$\tau_{V, \rm{tot}}$} & 22.4$_{-1.3}^{+1.2}$& 20.3$_{-2.3}^{+2.8}$ & 15.0$_{-0.7}^{+0.8}$ & 100$^{+5}_{-5}$\\

{$a_{sil}$} & 0.5 (fixed) & 0.5 (fixed) & 0.57$^{+0.05}_{-0.05}$ & 0.40$^{+0.05}_{-0.05}$ \\

{T$_{\rm{mol}}$} [K]   & 670$_{-30}^{+30}$ &  660$_{-30}^{+110}$ & -- & --\\
log(N$_{\rm{H_{2}O}}/\rm{cm}^{-2}$) &  22.0$_{-0.1}^{+0.1}$ & 22.5$_{-1.2}^{+0.2}$ & -- & --\\

log(N$_{\rm{CO}}/\rm{cm}^{-2}$) & 21.1$_{-1.7}^{+0.7}$ & -- & -- & --\\

log(N$_{\rm{SiO}}/\rm{cm}^{-2}$) & 19.9$_{-0.3}^{+0.3}$ & 19.8$_{-0.5}^{+0.5}$ & -- & --\\

{R$_{\rm{mol}}/10^{15}\rm{cm}$} &  2.1$_{-0.6}^{+0.2}$ & 2.7$^{+0.7}_{-0.7}$ & -- & -- \\

R$_{\rm{dust,in}}/10^{15}\rm{cm}$ & 1.3$_{-0.2}^{+0.2}$ & 3.6$^{+0.5}_{-0.5}$ & 2.1$_{-0.2}^{+0.2}$ & 2.7$_{-0.1}^{+0.1}$ \\ 

M$_{\rm{dust}}/10^{-5}\rm{M}_{\odot}$ & 4.2$_{-0.7}^{+0.7}$ & 30$^{+5}_{-5}$ & 7.5$^{+1.0}_{-1.0}$ & 77$^{+5}_{-5}$ \\

L$_{\rm{tot}}$/10$^{5}$L$_{\odot}$ & 3.3$_{-0.2}^{+0.2}$ & 4.2$_{-0.2}^{+0.2}$ & 1.48$_{-0.06}^{+0.06}$ & 4.62$^{+0.07}_{-0.05}\times10^{-2}$\\
\hline
\hline
\end{tabular}

\end{minipage}
\end{center}
\end{table*}
\endgroup

For AT\,2021blu, we find that the SED can be reproduced with a central star having a temperature $\approx2200$\,K, surrounded by a shell of warm dust with T$_{\rm{dust}}\approx1100$\,K, $\tau_{V}\approx22$, and an outer radius of $\approx78$\,A.U. The molecular shell has a colder temperature of $\approx670$\,K, and extends to a radius of $\approx170$\,A.U. 

In contrast, for AT\,2021biy, we find the central star has a similar temperature, but the dust shell is colder and larger, with T$_{\rm{dust}}\approx700$\,K, $\tau_{V}\approx21$ and an outer radius of $\approx200$\,A.U. The temperature of the molecular shell is $\approx 650$\,K, similar to AT\,2021blu, but the radius is much larger, $\approx 230$\,A.U. The column densities of the molecules are similar to AT\,2021blu.

\subsubsection{AT\,2018bwo and M31-LRN-2015}
\label{sec:bwo_m31_model_fitting}
As we do not have any spectroscopic data for AT\,2018bwo and M31-LRN-2015, we do not attempt to constrain their molecular content. Instead, we model their SEDs assuming only a central star surrounded by a dust shell using \texttt{DUSTY}. These sources were observed $\approx 7$ and 10 years since their eruptions respectively. At these late phases, silicate dust features are expected to dominate the molecular features in the 8--10\,\um region (see e.g. V838\,Mon, \citealt{Woodward2021}), so adding molecules to our models is not expected to substantially change the inferred dust properties. M31-LRN-2015 in particular shows a very deep silicate absorption feature in its SED, and so we are confident that modeling this absorption will still recover the true dust properties despite excluding molecules. We discuss the possible effects of molecular contributions to AT\,2018bwo in the next subsection. 

\begin{figure}
    \centering
    \includegraphics[width=0.5\textwidth]{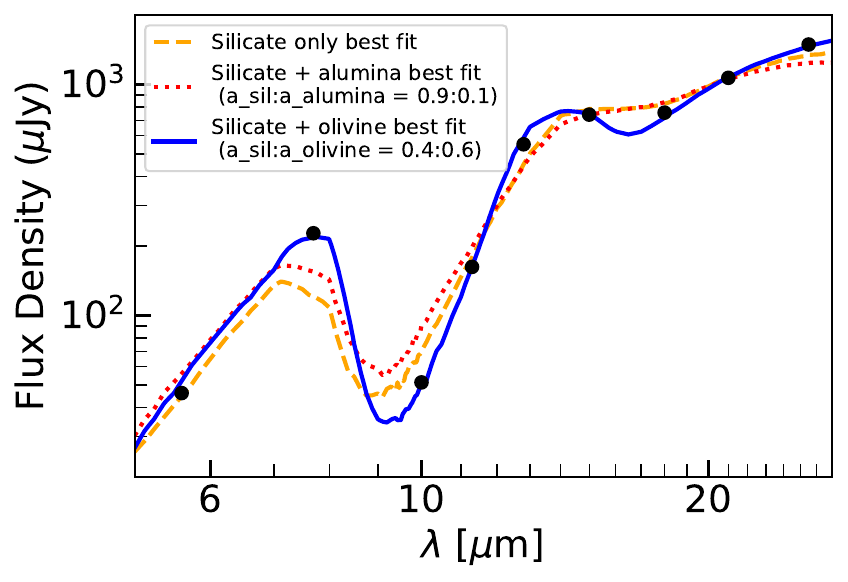}
    \caption{Best-fit models for M31-LRN-2015 with three different dust-compositions : pure siicate, silicate and alumina, and silicate and olivines. The siicate and olivine model fits the data the best.}
    \label{fig:m31_dust_composition}
\end{figure}

We model the multiband fluxes of these two sources using \texttt{DUSTY} with four free parameters (T$_{\rm{star}}$, T$_{\rm{dust}}$, $\tau_{V}$, and $a_{sil}$), where a$_{sil}$ is the relative abundance fraction of silicate dust. We assume similar dust grain sizes and density profiles as described in Section \ref{sec:blu_biy_model_fitting}. For AT\,2018bwo, we assume the same alumina and silicate dust composition but the ratio of silicate ($a_{sil}$) is a free parameter. For M31-LRN-2015, we did not find satisfactory fits using silicate and alumina dust. Instead, we find good matches with silicates and glassy olivines, as illustrated in Figure \ref{fig:m31_dust_composition}. This resembles the dust composition inferred for the Galactic merger OGLE-2002-BLG-360 \citep{Steinmetz2025}. Based on this, we assume that the dust composition for M31-LRN-2015 a mixture of silicates and glassy olivines, with their relative ratio as a free parameter. We use \texttt{emcee} to fit the models to the data and derive posterior distributions on the parameters. The derived parameters are listed in Table \ref{tab:mcmc_results}. The best-fit models are plotted in Figure \ref{fig:best_fit_models}. The posteriors are shown in Figures \ref{fig:corner_at2018bwo} and \ref{fig:corner_m31lrn2015} in the Appendix.

For AT\,2018bwo, we find a central star temperature of $\approx2600$\,K with a dust shell having T$_{dust}\approx650$\,K and $\tau_{V}\approx17$ and $a_{sil}=0.57^{+0.05}_{-0.05}$, and an outer radius of $\approx 380$\,A.U. In contrast, for M31-LRN-2015, we find a much cooler dust shell with T$_{\rm{dust}}\approx 320$\,K and a much larger $\tau_{V}\approx100$. The best-fit silicate fraction $a_{sil}=0.40^{+0.05}_{-0.05}$, and the outer radius of the dust shell is $\approx380$\,A.U.

\subsubsection{Caveats}
First, we note that AT\,2021blu and AT\,2021biy do not show any pronounced silicate features due to molecules, so their dust masses could be overestimated by our modeling. To test this, we attempted to fit their SEDs using models comprised of a blackbody stellar photosphere surrounded by molecules, without any dust. We find that while these models fit the 5-8\,\um region reasonably well, they underestimate the continuum flux at longer wavelengths. The characteristic continuum shape produced by the \texttt{DUSTY} models is required to reproduce these longer wavelength observations. Furthermore, for the best fitting no-dust models, the underlying blackbody photosphere has a temperature of $\approx900$\,K. No stars have such low photospheric temperatures, suggesting that the underlying photosphere is indeed formed by a warm dust shell. Through our MCMC modeling, we are able to constrain the properties of this dust shell reasonably well, suggesting that the reported dust masses are unlikely to be substantially overestimated, but future observations when more dust has formed and the silicate feature dominates the SED will verify the masses presented here for these two sources.

Next, the dust-masses of AT\,2018bwo and M31-LRN-2015 could be affected by molecules which are not included in the modeling for these two sources, as only photometric data are available for them. The effect of molecules is likely very small for M31-LRN-2015, as the dust properties are very well constrained by the strong silicate absorption feature that dominates its SED. For AT\,2018bwo, the effect of molecules could be higher. To test this, we converted the \emph{JWST} observations of AT\,2021blu to multiband photometric measurements by performing synthetic photometry over the spectrum, accounting for the \emph{JWST} filter transmission profiles, and fit it with \texttt{DUSTY} models without any molecules. We find a dust mass of $\approx5\times10^{-5}$\,M$_{\odot}$, only slightly larger than the value of $4.2\times10^{-5}$\,M$_{\odot}$ derived from dust and molecular modeling. Thus, we conclude that while the dust masses for AT\,2018bwo reported here could be overestimated, they are unlikely to be substantially higher than the true value.

Finally, for our modeling, we have assumed a simple spherically symmetric geometry for dust, surrounded by molecules. In reality, the geometry is likely more complicated, with dust and molecules mixed with each other. Sub-mm observations have revealed the complex bipolar morphology of this ejected material around the Galactic merger remnants \citep{Kaminski2018, Kaminski2021}. Although our simple models provide reasonably good fits to the SEDs, future studies that conduct 3D radiative transfer simulations can provide more information about the geometry of the ejecta.
\section{Discussion}
\label{sec:discussion}
\subsection{Dust masses and contribution to the cosmic dust budget}
\label{sec:dust_contribution}
\begin{figure*}
    \centering
    \includegraphics[width=0.8\textwidth]{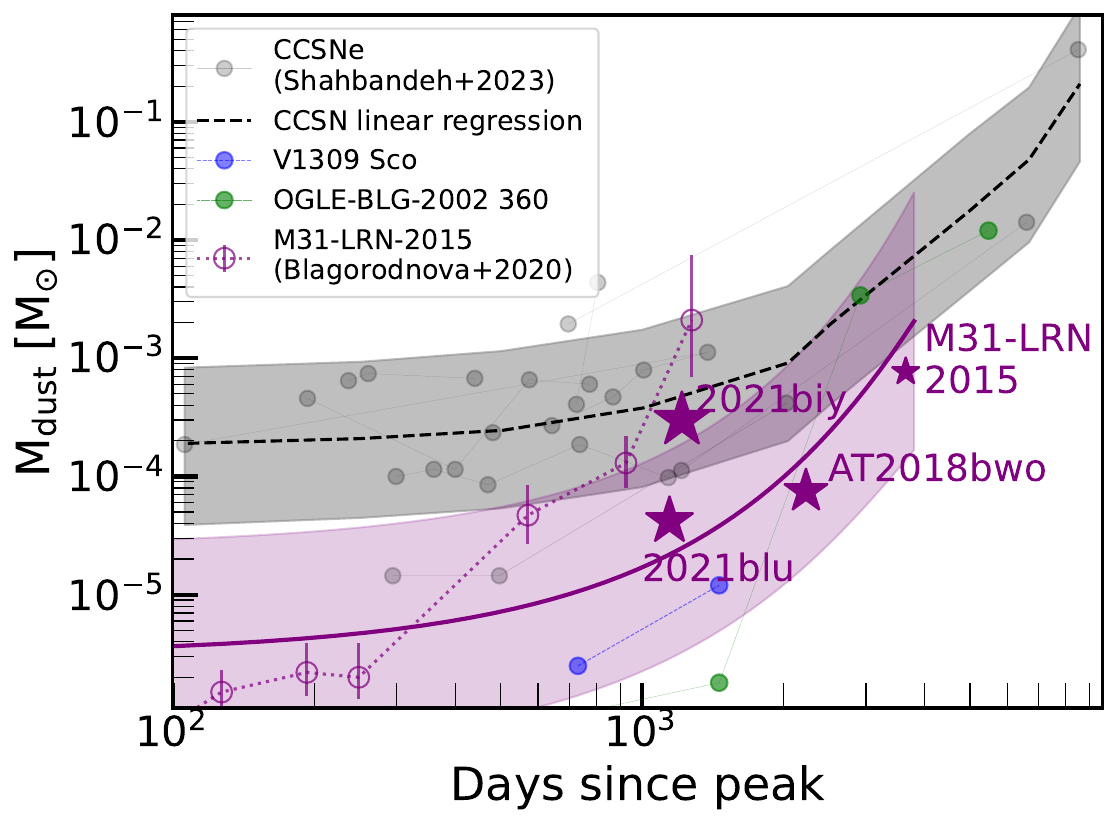}
    \caption{Dust masses inferred for the four LRNe in this paper (purple stars), compared with CCSNe (black dots), previous NIR-based measurements for M31-LRN-2015 (hollow purple circles), and Galactic LRNe V1309\,Sco (blue) and OGLE-2002-BLG-360 (green) with regression fits to LRNe (purple solid line) and CCSNe (black dashed line). The sizes of the purple stars indicate the progenitor masses of the four LRNe, ranging from $\approx$3-5\,M$_{\odot}$ for M31-LRN-2015 to $\approx$17-24\,M$_{\odot}$ for AT\,2021biy. The dust masses of the LRNe AT\,2021blu, AT\,2021biy, AT\,2018bwo, and M31-LRN-2015 are $\approx10$\%, 60\%, 6\% and 12\% of the median expected dust masses in CCSNe at similar phases. After accounting for occurrence rates relative to CCSNe, we estimate that the total dust produced by these LRNe is $\sim25\%$ of CCSNe. }
    \label{fig:dust_mass_comparison}
\end{figure*}

We use the dust shell parameters derived from \texttt{DUSTY} modeling to determine the mass of the dust formed around the LRNe using the following relation derived by \citet{Lau2025} ---

\begin{equation}
    \frac{M_{d}}{M_\odot} \approx 2.4 \times10^{-11} \left(\frac{R_\mathrm{in}}{50\, R_\odot}\right)^2 \left(\frac{\tau_V}{0.7}\right) \left(\frac{Y}{5}\right) \left(\frac{\kappa_V^{d}}{10^4\,\mathrm{cm}^2\,\mathrm{g}^{-1}}\right)^{-1}. 
\end{equation}

Using the values reported in Table \ref{tab:mcmc_results} and $\kappa_{V}^{d}\approx10^{4}$, the dust masses measured for the four LRNe are $\approx$4.2$\times10^{-5}$, 3$\times10^{-4}$, 7.5$\times10^{-5}$, and 7.7$\times10^{-4}$\,M$_{\odot}$ for AT\,2021blu, AT\,2021biy, AT\,2018bwo, and M31-LRN-2015 respectively. The exact values and uncertainties are reported in Table \ref{tab:mcmc_results}. Of these four sources, dust masses have previously been reported only for M31-LRN-2015 at phases of 43 to 1276 days since maximum \citep{Blagorodnova2020}. The dust masses measured in their study increased from $\approx10^{-7}$\,M$_{\odot}$ at 46 days to $\sim10^{-3}$\,M$_{\odot}$ at 1276 days during this period. This does not agree with our late-time measurement of $7.7\times10^{-4}$\,M$_{\odot}$ at +3647 days. However, the SEDs modeled in that study contained data only at wavelengths shorter than 5\um, with the final epoch (1276 days) SED comprising only one detection at 4.5\um and three non-detections at shorter wavelengths. Consequently, some of their measurements are likely  inaccurate, and at least the final-epoch measurement is ruled out by our longer wavelength observations. 

Figure \ref{fig:dust_mass_comparison} shows our dust mass measurements of the four LRNe as a function of the time since their eruption (purple stars). We also include the previous measurements of M31-LRN-2015 for comparison as hollow circles. In general, the dust masses appear to increase with phase since the eruption. To visualize this, we fit a gaussian process regression model to the LRN measurements and plot the range of expected dust masses as a purple shaded region in Figure \ref{fig:dust_mass_comparison}. For this fit, we included all four of our measurements and all but the last two epochs of the previous M31-LRN-2015 measurements. Also plotted in Figure \ref{fig:dust_mass_comparison} are the dust-masses for core collapse supernovae (CCSNe, taken from \citealt{Shahbandeh2023}) and a regression fit to the CCSN dust masses (also from \citealt{Shahbandeh2023}). Comparing our \emph{JWST} measurements to CCSNe, we find that the dust masses for AT\,2021blu, AT\,2021biy, AT\,2018bwo, and M31-LRN-2015 are $\approx10$\%, 60\%, 6\% and 12\% of the median expected dust masses in CCSNe at similar phases. The earlier ($<500$\,day) M31-LRN-2015 measurements from \citet{Blagorodnova2020} suggest lower dust masses ($\approx2\%$) compared to CCSNe at phases $<500$ days. AT\,2021biy has a larger dust mass than AT\,2021blu at a similar phase. This is likely due to a combination of a larger ejected mass from the more massive progenitor of AT\,2021biy, and dust-formation in the outer layers of the ejecta due to shock-interactions with CSM. 

To estimate the contribution of LRNe to the cosmic-dust budget relative to CCSNe, we convolve the dust masses with the LRN-volumetric rates. We use the rates and luminosity function derived in \citet{Karambelkar2023} (which is valid for M$_{\rm{r, peak}}<-11$) for AT\,2018bwo, AT\,2021blu, and AT\,2021biy. For M31-LRN-2015 (M$_{\rm{r, peak}}=-9.5$), we use the Galactic merger rate and luminosity function from \citet{Kochanek14_mergers}, which predicts a rate of $\sim0.03$\,yr$^{-1}$ for mergers with M$_{V}\approx-10$. We assume a local volumetric CCSN rate of $\approx$10$^{-4}$\,Mpc$^{-3}$\,yr$^{-1}$ \citep{Perley2020} and a Galactic CCSN rate of $\approx0.016$\,yr$^{-1}$ \citep{Rozwadowska2021} to convert LRN rates to those relative to CCSNe. We find that the rate-corrected total dust mass in these four LRNe is $\sim25\%$ of the median dust mass in CCSNe, dominated by M31-LRN-2015 which is the least luminous and hence the most common of the four LRNe. We note that if we extrapolate the \citet{Karambelkar2023} luminosity function to the luminosity of M31-LRN-2015, we obtain a larger dust mass of $\approx40$\% of CCSNe. 

To summarize, we find that the dust contribution of LRNe to the cosmic dust-budget is $\approx25\%$ of CCSNe. However, this is likely a lower limit due to the reasons stated below.

First, for all four LRNe, dust formation probably continues beyond the epochs probed by the \emph{JWST} observations. All four LRNe have different progenitor masses and luminosities, and likely have different dust-formation timescales. For example, we find the largest dust mass for M31-LRN-2015 --- the source with the lowest luminosity and the lowest progenitor mass in our sample. AT\,2021blu, and AT\,2021biy have progenitors $\sim$4-7 times more massive than M31-LRN-2015 and their luminosities suggest about an order of magnitude more ejected mass than M31-LRN-2015. Thus, they are expected to produce more dust in the long term than M31-LRN-2015, which would bring their terminal dust yields closer to CCSNe. However, the presence of a more massive remnant could keep the ejecta warm for longer periods, prolonging the dust formation timescales compared to M31-LRN-2015. Additionally, the more massive progenitors could eventually explode as core-collapse supernovae. Depending on the delay between the merger and the supernova, some of the dust produced during the merger could potentially be destroyed by the supernova explosion. Long-term IR monitoring of a wider variety of LRNe will help determine their terminal dust yields.

Second, while CCSNe produce large amounts of dust, a substantial fraction of this dust can be destroyed by the reverse shock passing through the supernova ejecta. Several theoretical studies have attempted to estimate the efficiency of dust destruction in SNe and found results ranging from little to complete destruction, depending on the grain size and dust composition. In general, smaller grains are easily destroyed, while larger grains can survive longer \citep{Nozawa2007}. For silicate dust, several studies have found dust survival fractions of $\sim$10--20\% \citep{Slavin2020, Micelotta2016, Silvia2010, Bianchi2007}. Lower survival fractions of a few percent are predicted by \citet{Bocchio2016, Kirchschlager2019, Priestley2022}, and higher survival fractions are supported by \citet{Nozawa2007} (20-100\%) and \citet{Nath2008} (>80\%). \citet{Biscaro2016} find that dust survival also depends on the SN type with most of the dust destroyed in Type IIb SNe, 14-45\% dust surviving in Type II-P SNe, 6--11\% survival in CasA, and 42--98\% survival in dense ejecta of SN\,1987A. 

While the dust destruction in LRNe has not been quantified, their substantially lower velocities ($\sim100$\,km\,s$^{-1}$) compared to CCSNe ($\sim1000$\,km\,s$^{-1}$) suggest lesser dust destruction. The rate of dust destruction is expected to increase steeply with shock velocity (see e.g. \citealt{McKee1989, Jones1994}). \citet{Slavin2020} find that the surviving fraction of dust depends on the relative velocity of the dust grains and the shocked gas, with a dust survival rate of more than 90\% for a relative velocity of 100\,km\,s$^{-1}$, and less than 40\% for a relative velocity of 1000\,km\,s$^{-1}$. Thus, a large fraction of the dust produced in LRNe could eventually make its way into the ISM, unlike CCSNe. A dust survival rate of 10--20\% for CCSNe would then imply comparable contributions from LRNe and CCSNe to the cosmic dust budget. Future studies quantifying the destruction of dust by low-velocity shocks in LRNe will help to address this. 

Finally, our four LRNe do not capture the full variety of stellar mergers. The progenitors of all four LRNe had YSG primaries, but stellar mergers with red giant or supergiant primaries are expected to produce more dust \citep{Macleod2022}. In Figure \ref{fig:dust_mass_comparison}, we also show the inferred dust masses for the Galactic stellar merger OGLE-BLG-2002-360 -- the only known example of a giant star merger (taken from \citealt{Tylenda13, Steinmetz2025}). About fifteen years since its eruption, this source has formed $\approx10^{-2}$\,M$_{\odot}$ of dust, similar to CCSNe at this phase. Historically, optical surveys have missed these dusty mergers but this population is being unveiled by the advent of infrared time-domain surveys \citep{Jencson2019, Karambelkar2025}. Future IR observations of the wider variety of stellar mergers will be crucial to quantify the full dust contributions of stellar mergers to the cosmic dust budget.

In conclusion, our observations suggest that the dust masses produced in LRNe are $\gtrsim25$\% of those in CCSNe. However, the higher dust destruction fraction in CCSNe compared to LRNe together with possibly longer timescales of dust formation in massive stellar mergers could make the dust contributions of LRNe comparable to CCSNe. Our results strongly advocate for multi-epoch IR monitoring of a wider variety of LRNe to trace their dust formation timescales and terminal dust yields. 

\subsection{Water vapor and molecules}
The \emph{JWST} spectra of AT\,2021blu and AT\,2021biy are dominated by features of water vapor and other O-rich molecules. This is not surprising, as large molecular clouds have been detected around the Galactic stellar mergers \citep{Lynch04, Lynch2007, Kaminski2015AA, Kaminski2018, Kaminski2021}, and NIR spectra of several LRNe show absorption features due to these molecules \citep{Karambelkar2023, Blagorodnova2021, Pastorello2022, Cai2022}. We used our \emph{JWST} observations to estimate the column densities of these molecules for the two extragalactic LRNe. The derived molecular column densities from Table \ref{tab:mcmc_results} for water vapor, CO, and SiO of $\approx10^{22}$, $10^{21}$, and $10^{20}$\,cm$^{-2}$ are similar to the values measured for V838\,Mon \citep{Lynch04}. 

The water vapor column densities in LRNe are generally higher than those measured in red giant and supergiant stars, which have column densities of $\sim10^{18}$\,cm$^{-2}$ \citep{Tsuji2002}. Some supergiants, such as Betelgeuse and $\mu$Cep, have been identified with possible molecular spheres outside of their photospheres (a.k.a. \emph{molspheres} \citealp{Tsuji2006}) with water vapor column densities of $\approx10^{20}$\,cm$^{-2}$ \citep{Tsuji2000, Ohnaka2004, Tsuji2006} -- still lower than the values found for LRNe. The molecular shell radii measured for our LRNe suggest expansion velocities of $\sim250$ and 300\,km\,s$^{-1}$ for AT\,2021blu and AT\,2021biy -- consistent with the molecules being formed in the outer layers of expanding ejecta. The enhanced column densities in the LRNe thus suggest large amounts of water vapor streaming away from the merger remnant. Using the radii, we estimate the mass of the water vapor to be $\sim1-3\times10^{-3}$\,M$_{\odot}$ for AT\,2021blu and $10^{-3}-10^{-2}$\,M$_{\odot}$ for AT\,2021biy. 

Our observation of large amounts of water vapor streaming away from stellar merger remnants supports the recent association of Galactic water-fountain sources with common-envelope events \citep{Khouri2022}. These water fountains are characterized by $>100$\,km\,s$^{-1}$ water maser emission arising from jets excavating dusty circumstellar envelopes, and have unusually high mass loss rates that can be explained by a recent common-envelope event. They also have a characteristic geometry with bipolar lobes around the remnant surrounded by slow expanding gas in a toroidal configuration perpendicular to the lobes \citep{Sahai2017}. Future 3D-modeling of our observations of AT\,2021blu and AT\,2021biy is required to test whether such a geometry is present around their remnants. 

Finally, the detection of molecules in the mid-IR spectra of these massive stellar mergers provides an opportunity to study the long-term evolution of molecules in oxygen-rich environments. It will be particularly interesting to study the transition of water-vapor to water-ice, which has been detected in the late-time IR spectrum of the Galactic stellar merger V4332\,Sgr \citep{Banerjee2004} through its broad absorption feature at 3\um. Continued spectroscopic observations of the 8--10\um range that covers the silicate-dust feature will enable studying the dust-condensation sequence of oxygen-rich environments, as more dust condenses onto the molecular seeds (e.g. \citealt{Verhoelst2009}).

\subsection{Late-time evolution of LRNe}
\label{sec:late_time_evol}
\begin{figure}
    \centering
    \includegraphics[width=0.5\textwidth]{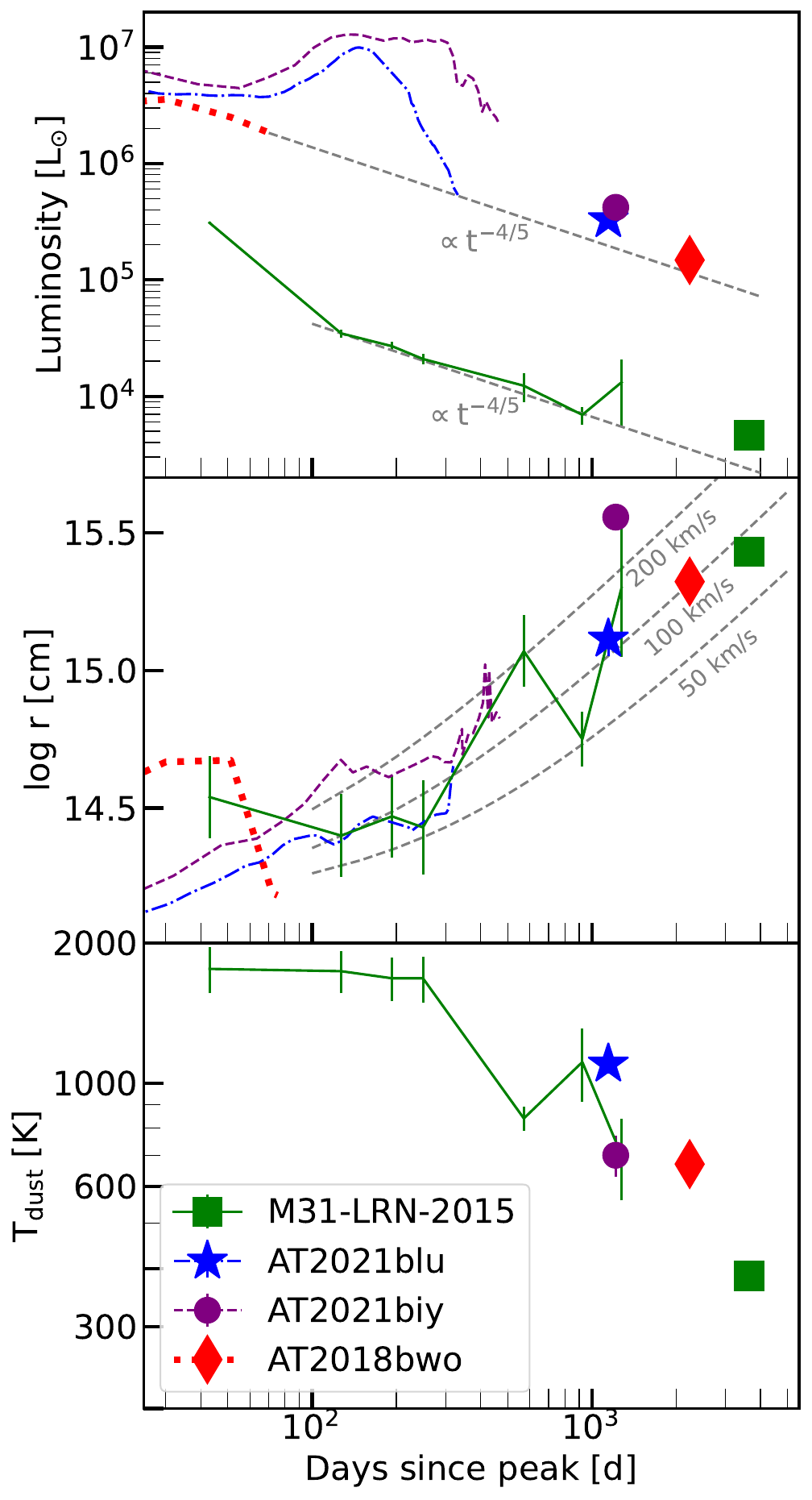}
    \caption{Evolution of LRN properties. \emph{JWST} measurements are indicated with stars, while solid lines show previous measurements wherever available. \emph{Top:} The \emph{JWST} luminosities of all four LRNe are higher than that extrapolated from their previous trends. The luminosities are inconsistent with  gravitational contraction of an inflated remnant ($\propto$t$^{-4/5}$), and suggest shock-interactions with CSM. \emph{Middle:} The dust shell inner radii are consistent with expansion velocities of 100-250\,km\,s$^{-1}$, with the highest velocity for AT\,2021biy that has the most massive progenitor of the four LRNe. AT\,2021biy also has a colder dust shell temperature than AT\,2021blu at a similar phase \emph{(Bottom panel)}.}
    \label{fig:lrn_properties_evolution}
\end{figure}
We now use our \emph{JWST} observations to study the late-time evolution of LRN remnants. Figure \ref{fig:lrn_properties_evolution} shows the evolution of the luminosities, dust-shell radii, stellar and dust temperatures for the four LRNe. In addition to our \emph{JWST} measurements, the previous values of these parameters for these sources are taken from \citet{Blagorodnova2020} and \citet{Macleod2022} for M31-LRN-2015, \citet{Cai2021} and \citet{Karambelkar2023} for AT\,2021biy, \citet{Pastorello2022} for AT\,2021blu, and \citet{Blagorodnova2021} for AT\,2018bwo. 

The \emph{JWST} luminosities show that all four LRNe have faded since their previous observations (Figure \ref{fig:lrn_properties_evolution} top panel). However, all four are brighter than extrapolations from their previous trends, suggesting that the luminosity decline has slowed down. This has been observed previously in V838\,Mon, V4332\,Sgr, and M31-LRN-2015, whose luminosities at late times declined as $L \propto t^{-4/5}$, consistent with the expectation for a gravitationally contracting inflated envelope around the remnant \citep{Tylenda05a, Tylenda05b, Blagorodnova2020}. For M31-LRN-2015, this trend was observed for 2.5 years after the peak, following which its luminosity appeared to increase, possibly due to shocks \citep{Blagorodnova2020}. Consistent with this, our \emph{JWST} luminosity of M31-LRN-2015 is $\approx3$ times higher than the expectation from the $t^{-4/5}$ trend. Similarly, the other three LRNe have higher luminosities than that expected from gravitational contraction, suggesting that their shocks could be contributing to their late-time luminosities. We note that all LRNe are more luminous than their progenitors at the time of the \emph{JWST} observations. 

The second and bottom panel of Figure \ref{fig:lrn_properties_evolution} shows the evolution of the inner radii and temperatures of the dust shells respectively. The radii of AT\,2021blu, AT\,2018bwo, and M31-LRN-2015 are consistent with expansion speeds of $\approx100$\,km\,s$^{-1}$, while AT\,2021biy has a larger expansion velocity of $\,250$\,km\,s$^{-1}$. The dust temperatures generally decline with time. We note that while AT\,2021blu and AT\,2021biy were observed at similar phases of $\approx1200$ days, AT\,2021biy has a dust-shell that is three times larger,  $\approx$400\,K cooler (Figure \ref{fig:lrn_properties_evolution}, bottom panel), and eight times more massive than AT\,2021blu. AT\,2021biy also had the longest lightcurve plateau of all LRNe, showed rebrightenings after the plateau ended likely powered by shocks, is the most luminous of the four LRNe in our sample, and has the most massive progenitor of the four \citep{Cai2022}. These properties suggest a larger ejected mass in AT\,2021biy than the other LRNe, and also point to stronger CSM interaction that can produce large amounts of dust in the outer layers of the ejecta due to shocks. 

\subsection{Dust-to-gas ratios in LRNe}
\begin{figure}
    \centering
    \includegraphics[width=\linewidth]{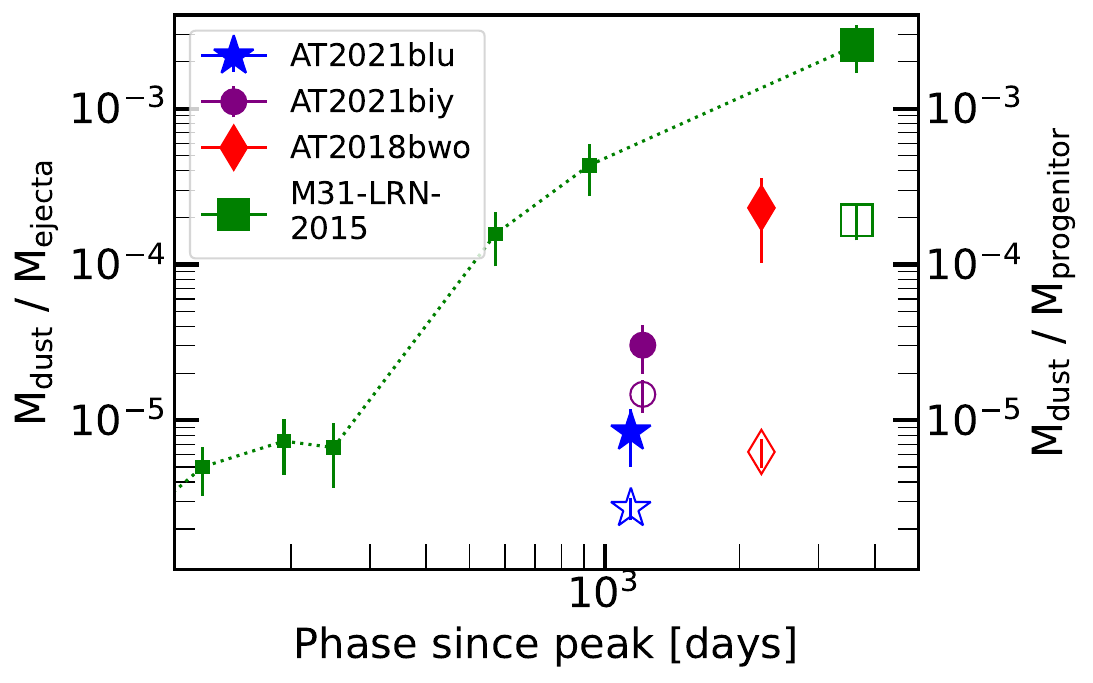}
    \caption{Dust-to-gas ratios (i.e. ratios of dust masses to the total ejected masses, solid symbols) as a function of days since peak for the four LRNe presented in this paper. The dust-to-gas ratios in M31-LRN-2015, the lowest mass merger, exceed those of the more massive mergers by an order of magnitude, suggesting that dust formation might occur on faster timescales in low-mass mergers. While the dust-to-gas ratios generally increase with time, the values even at $\sim$10 years since peak are lower than the value of $\approx$10$^{-2}$ seen in the interstellar medium, suggesting that the dust formation might continue beyond the timescales probed by our observations. Also indicated are the ratios of dust masses to the LRN progenitor masses (hollow symbols).}
    \label{fig:dust_gas_ratios}
\end{figure}

Figure \ref{fig:dust_gas_ratios} shows the dust-to-gas ratios calculated by dividing the dust masses by the total ejected masses for the four LRNe. We use the lightcurve-based estimates for the total ejected masses of $\approx$0.3\,M$_{\odot}$ for M31-LRN-2015 \citep{MacLeod17}, $\approx$0.15-0.5\,M$_{\odot}$ for AT2018bwo \citep{Blagorodnova2021}, $\sim$5\,M$_{\odot}$ for AT2021blu \citep{Pastorello2022}, and $\sim$10\,M$_{\odot}$ for AT2021biy \citep{Cai2022}. We note that the ejected mass estimates for AT2021blu and AT2021biy assume recombination powered lightcurves, which is likely an overestimate (see \citealt{Matsumoto2022}). Our \emph{JWST} dust mass measurements yield dust-to-gas ratios ranging from $\sim10^{-5}$ for AT2021blu and AT2021biy at about three years since peak to $\sim2\times10^{-3}$ for M31-LRN-2015 at ten years since peak. For M31-LRN-2015, we also plot dust-to-gas ratios for phases earlier than 1000 days since peak using the dust mass measurements from \citet{Blagorodnova2020}, which show that the ratios increased from 10$^{-5}$ to $10^{-3}$ in ten years. Figure \ref{fig:dust_gas_ratios} also illustrates the differences in dust-formation timescales for the LRNe in our sample.

First, Figure \ref{fig:dust_gas_ratios} shows that the dust-to-gas ratios in M31-LRN-2015, the lowest mass merger, exceed the more massive mergers at similar phases by about one order of magnitude. This suggests that dust formation might occur on faster timescales in low-mass mergers, consistent with the observation that lower-mass mergers exhibit a faster transition to red optical colors than their more massive counterparts, likely due to faster cooling of their ejecta \citep{Blagorodnova2021}. 

Second, if the dust-to-gas ratios for the more massive mergers eventually increase to $\approx10^{-3}$ similar to M31-LRN-2015, the total dust masses for AT2021blu and AT2021biy will be $\approx 5\times10^{-3}$\,M$_{\odot}$ and $\approx10^{-2}$\,M$_{\odot}$ --- comparable to some of the dustiest CCSNe in Figure \ref{fig:dust_mass_comparison}. This would increase the contribution of LRNe to the cosmic dust-budget, as mentioned in Section \ref{sec:dust_contribution}.

Third, it is possible that substantial dust formation continues well beyond the timescales probed by our observations. The dust-to-gas ratios, even for M31-LRN-2015 at ten years since peak, are lower than the value of 10$^{-2}$ seen in the interstellar medium, which is commonly assumed for LRNe and other transients (e.g., \citealt{Blagorodnova2020}). Simulations of binaries undergoing CEE show that dust formation continues in the inner regions of the ejecta for several years after the dynamical inspiral \citep{Iaconi2020}. \citet{Gonzalez-Bolivar2024} and \citet{Bermudez2024} simulated carbon dust formation in CE interactions of low mass AGB stars with a 0.6\,M$_{\odot}$ companion. They find that dust formation starts about 2-3 years into their simulations during the early inspiral phase, but really takes off ten years later when it reaches $\approx10^{-3}$\,M$_{\odot}$ and increases steadily to a plateau of $\approx10^{-2}$\,M$_{\odot}$ after 30-40 years. \citet{Iaconi2020} find that for a 0.88\,M$_{\odot}$ red-giant primary, the dust formation occurs between 300 -- 5000 days after the dynamical inspiral ends, reaching terminal values of $\sim10^{-3}$\,M$_{\odot}$ at 5000 days. While these simulated systems differ substantially from the LRNe studied in this paper, it seems reasonable that the dust masses in LRNe will continue to increase for several decades since eruption, beyond the maximum phase of ten years that is probed by our observations. 

Future observations of older LRNe will be crucial to trace the dust formation curve in LRNe, determine their terminal dust yields and dust-to-gas ratios. 

\section{Summary and way forward}
\label{sec:summary}
In this paper, we presented \emph{JWST} observations of four LRNe originating from extragalactic massive stellar mergers. The four LRNe spanned a wide progenitor primary mass range (3--24\,M$_{\odot}$) and a wide range of phases since merger (1100--3600\,days). We modeled the infrared SEDs of these LRNe to measure their dust and molecular content. We find -

\begin{itemize}[leftmargin=*]
    \item The dust masses in the four LRNe AT\,2021blu, AT\,2021biy, AT\,2018bwo, and M31-LRN-2015 are $\approx$4.2$\times10^{-5}$, 3$\times10^{-4}$, 7.5$\times10^{-5}$, and 7.7$\times10^{-4}$\,M$_{\odot}$. This corresponds to  $\approx10$\%, 60\%, 6\% and 12\% of the median expected dust masses in CCSNe at similar phases.
    \item After accounting for occurrence rates, the total dust mass produced in LRNe is estimated to be $\sim25\%$ of that produced in CCSNe. This is likely a lower limit, as dust-formation is expected to continue beyond the epochs coverred by our observations. Furthermore, a lower dust-destruction fraction in LRNe than CCSNe may make the dust contributions of LRNe comparable to CCSNe.
    \item We also find large column densities of water vapor, CO, and SiO around the merger remnant, making LRNe important laboratories to study the evolution of these molecules as they potentially transition to form ices with future observations. The detection of water vapor also supports the association of water fountain sources with CEE remnants. 
    \item The late-time luminosities of LRNe indicate that they might be undergoing shock-powered rebrightenings due to interaction with CSM. The \emph{JWST} luminosities of all four sources are higher than their pre-outburst progenitor luminosities.
    \item We estimate dust-to-gas ratios for the four LRNe, ranging from $\sim10^{-5}$ at 1144 days since peak for AT2021blu to $\sim2\times10^{-3}$ at 3647 days since peak for M31-LRN-2015. The ratios suggest that dust formation occurs on a faster timescale in low-mass mergers than their more massive counterparts. Dust formation might also continue for several years beyond the timescales probed by our observations. 
\end{itemize}

This work highlights the potential of LRNe in understanding cosmic dust sources, studying dust-formation and evolution of molecules, and  probing mass-loss during CEE phase. Future observations with \emph{JWST} of a broader sample of LRNe will be crucial to determine the terminal dust yields of LRNe, and study the evolution of the dust and molecular properties. These studies will set the stage for future IR missions such as the \emph{Nancy Grace Roman Space Telescope} and the concept mission PRobe far-Infrared Mission for Astrophysics (PRIMA).

\section*{Acknowledgements}
This work is based on observations made with the NASA/ESA/CSA James Webb Space Telescope. The data were obtained from the Mikulski Archive for Space Telescopes at the Space Telescope Science Institute, which is operated by the Association of Universities for Research in Astronomy, Inc., under NASA contract NAS 5-03127 for JWST. These observations are associated with program \#4244. Some of the data presented herein were obtained at Keck Observatory, which is a private 501(c)3 non-profit organization operated as a scientific partnership among the California Institute of Technology, the University of California, and the National Aeronautics and Space Administration. The Observatory was made possible by the generous financial support of the W. M. Keck Foundation. The authors wish to recognize and acknowledge the very significant cultural role and reverence that the summit of Maunakea has always had within the Native Hawaiian community. We are most fortunate to have the opportunity to conduct observations from this mountain. N. B. acknowledges to be funded by the European Union (ERC, CET-3PO, 101042610). Views and opinions expressed are however those of the author(s) only and do not necessarily reflect those of the European Union or the European Research Council Executive Agency.

\begin{appendix}
\counterwithin{figure}{section}
    \section{Modeling the dust and molecular emission}
    As discussed in Section \ref{sec:model_fitting}, we model the emission from a central star and dust using \texttt{DUSTY}, and then add a plane-parallel layer of molecules around it. We assume the outer radius of the star+dust shell configuration is R$_{*}$, and the molecular slab is characterized by its temperature (T$_{\rm{mol}}$), radius (R$_{\rm{mol}}$) and optical depth ($\tau_{\nu}$) with R$_{\rm{mol}}>R_{*}$. Under these conditions, the radiative transfer equation is 
    \begin{equation}
        I_{\nu} = I_{\nu_0}e^{-\tau_\nu} + B_{\nu}(T)(1 - e^{-\tau_\nu})
    \end{equation}

    and the emergent flux density (given that $I_{\nu_0} = 0$ for $R_{\rm{mol}}>r>R_{*}$) is given by 
    \begin{equation}
        f_{\nu} = \frac{\pi R_{*}^{2}}{D^2} I_{\nu_0}e^{-\tau_\nu} + \frac{\pi R_{\rm{mol}}^2}{D^2}(1 - e^{-\tau_\nu}) = f_{\nu}^{d}e^{-\tau_\nu} + \frac{\pi R_{\rm{mol}}^2}{D^2}B_{\nu}(T)(1 - e^{-\tau_\nu})
    \end{equation}

    where $f_{\nu}^d = \frac{\pi R_{*}^{2}}{D^2} I_{\nu_0}$ is the emergent flux density from the star+dust unaffected by the presence of molecules.\\
    \\
    Unfortunately, \texttt{DUSTY} outputs neither $f_{\nu}^{d}$ nor $I_{\nu_{0}}$, but instead outputs the shape of the spectrum i.e. normalized $\nu f_{\nu}$. From this, we can only determine a normalized flux density $f_{\nu}^{d,n}$ such that $\int f_{\nu}^{d,n} d\nu = 1$ and the required model flux density $f_{\nu}^{d} = Cf_{\nu}^{d,n}$. 

    \begin{equation}
        f_{\nu} = Cf_{\nu}^{d,n}e^{-\tau_\nu} + \frac{\pi R_{\rm{mol}}^2}{D^2}B_{\nu}(T)(1 - e^{-\tau_\nu}) = \frac{\pi R_{\rm{mol}}^2}{D^2} \times [C_1 f_{\nu}^{d,n}e^{-\tau_\nu} + B_{\nu}(T)(1 - e^{-\tau_\nu})]
    \end{equation}

In our fitting process, the free parameters are R$_{\rm{mol}}$, C$_{1}$ and $\tau_{\nu}$ (which incorporates the column densities as free parameters). Once C$_{1}$ and R$_{mol}$ are determined, the radius of the dust shell can be estimated using the r$_1$ parameter provided as part of \texttt{DUSTY} output. We first calculate the integrated bolometric luminosity of the \texttt{DUSTY} model alone 
    \begin{equation}
        L_{d} = \int 4\pi D^{2} f_{\nu}^{d} d\nu = \int 4\pi D^2 \frac{\pi R_{mol}^{2}}{D^2}C_1 f_{\nu}^{d,n} d\nu = 4\pi^2 R_{mol}^2 C_1 \int f_{\nu}^{d,n}d\nu = 4\pi^2 R_{mol}^2 C_1
    \end{equation}
    and then use this to calculate the radius 
    \begin{equation}
    R_{*} = r_{1} (\frac{L_d}{10^4 L_{\odot}})^{\frac{1}{2}}Y  = r_{1} (\frac{4\pi^2 R_{mol}^2 C_1}{10^4 L_{\odot}})^{\frac{1}{2}}Y   
    \end{equation}
     where Y is the shell-thickness.

\begin{figure*}[hbt]
    \centering
    \includegraphics[width=0.9\textwidth]{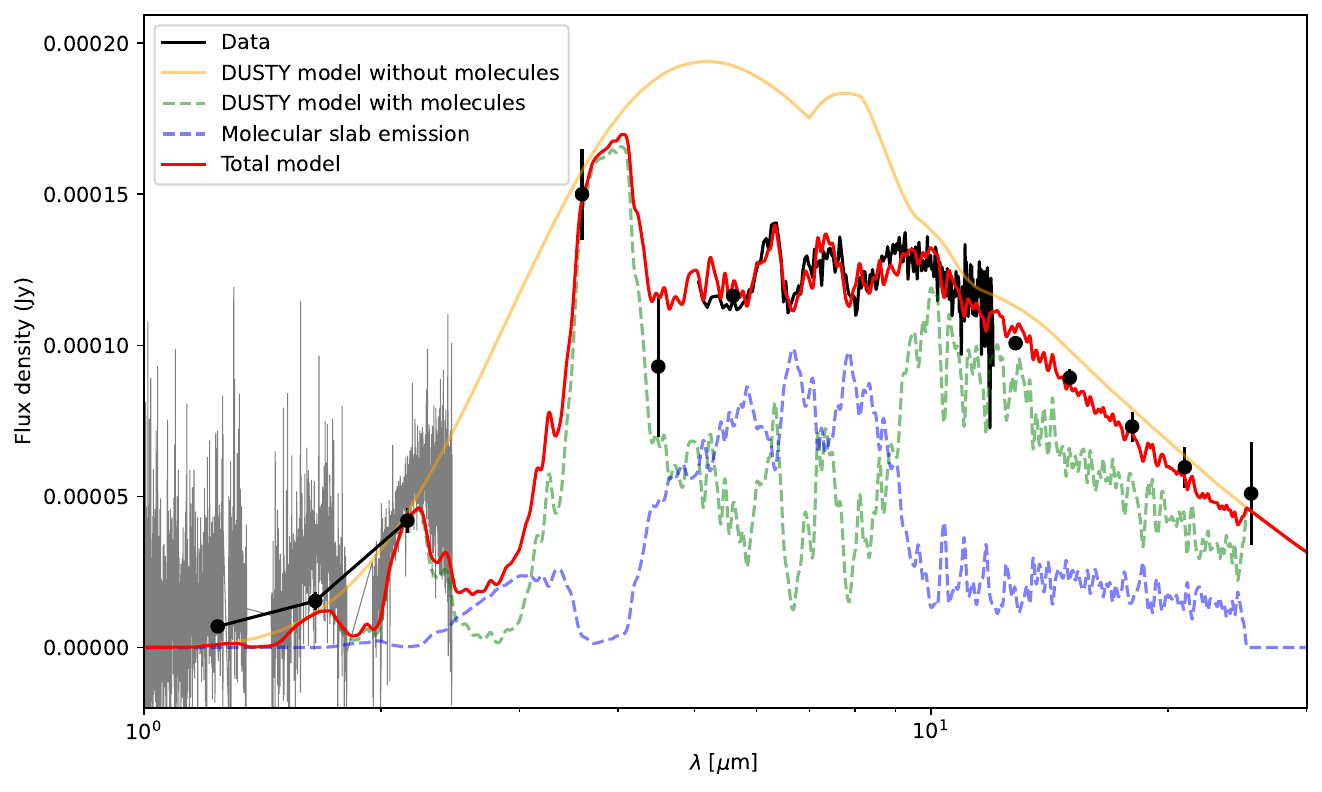}
    \caption{A breakdown of the different model components for AT\,2021blu: \texttt{DUSTY} model (blackbody star + dust shell, orange), \texttt{DUSTY} model attenuated with molecules (dashed green line), emission from molecular shell (dashed blue line), total model (red line = green + blue) that fit to the data (black).}
    \label{fig:at2021blu_component_fits} 
\end{figure*}

An example of the different model components is shown in Figure \ref{fig:at2021blu_component_fits}.

\begin{figure}
    \centering
    \includegraphics[width=0.8\textwidth]{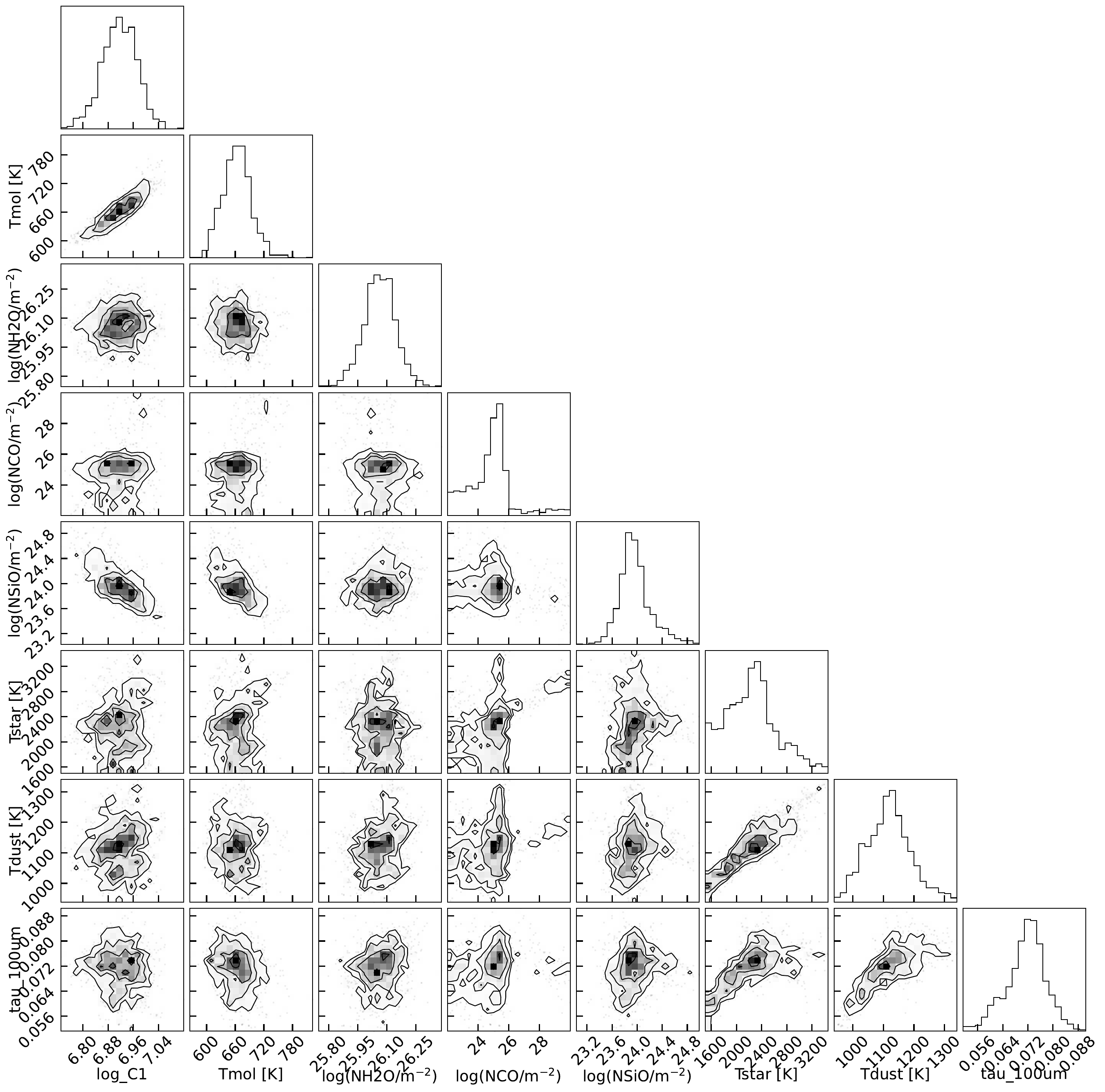}
    \caption{Posterior probability distributions of parameters for AT\,2021blu. Note that \texttt{tau\_{100um}} represents the dust optical depth at 100\,\um. }
    \label{fig:corner_at2021blu}
\end{figure}

\begin{figure}
    \centering
    \includegraphics[width=0.8\textwidth]{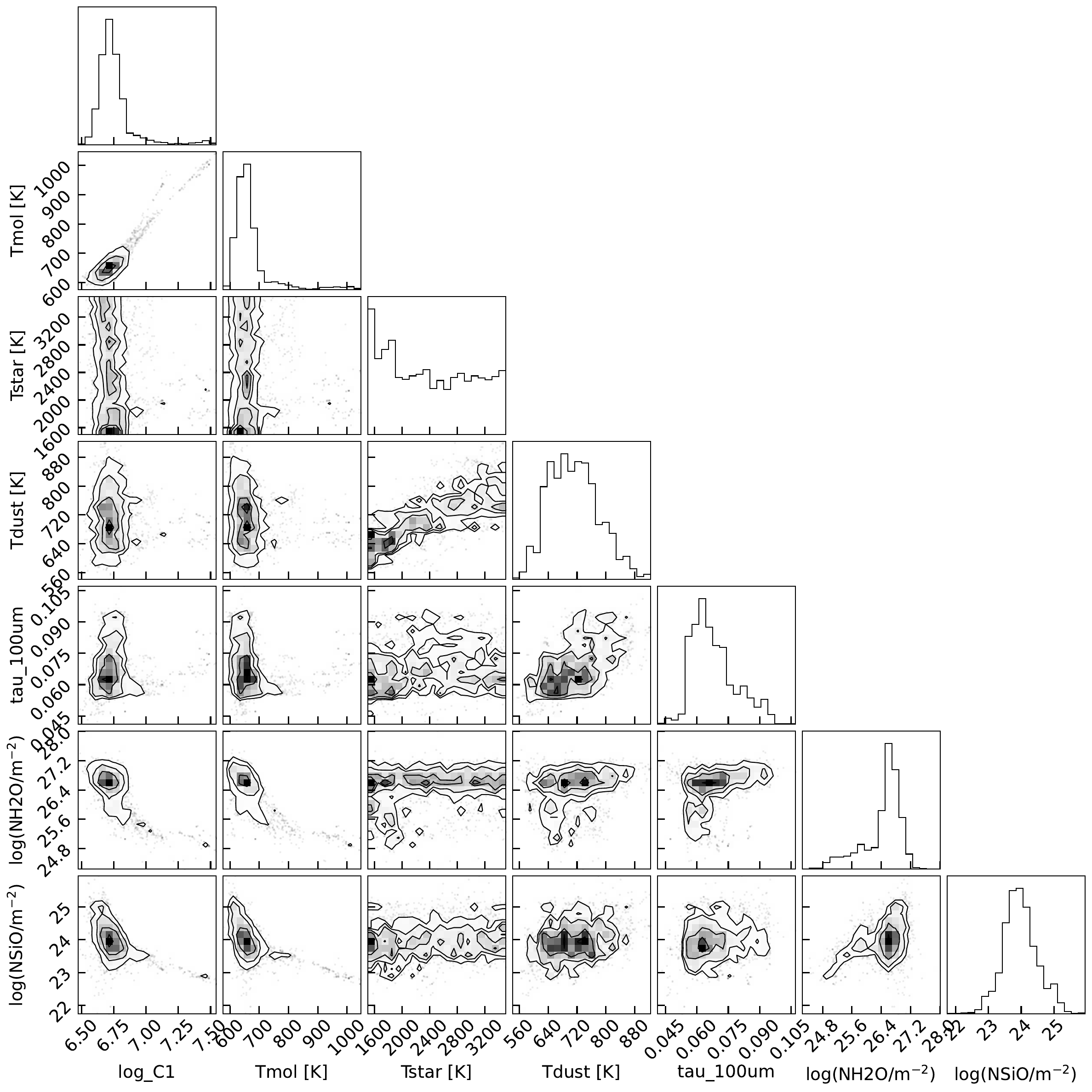}
    \caption{Posterior probability distributions of parameters for AT\,2021biy. Note that \texttt{tau\_{100um}} represents the dust optical depth at 100\,\um.}
    \label{fig:corner_at2021biy}
\end{figure}

\begin{figure}
    \centering
    \includegraphics[width=0.8\textwidth]{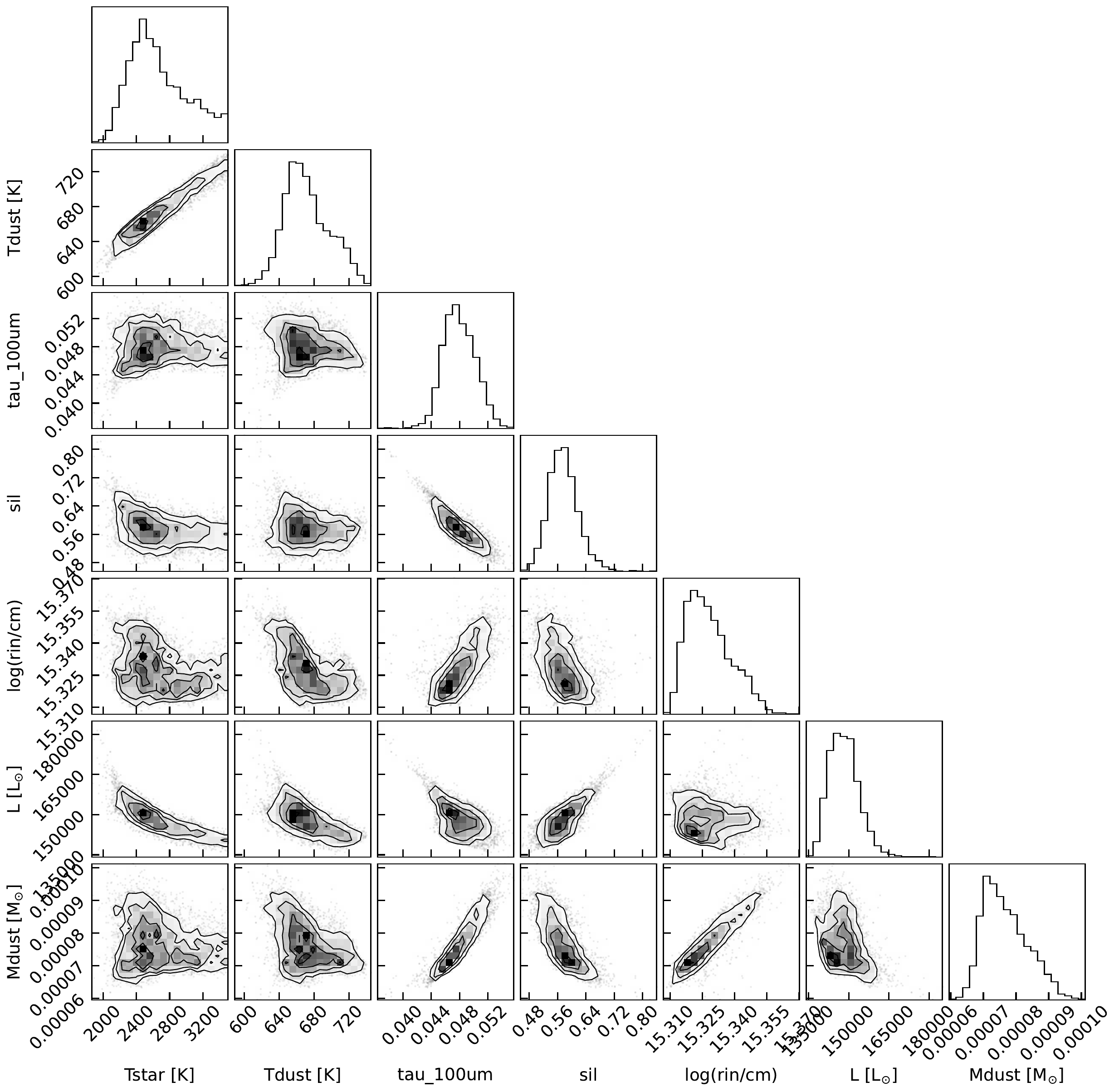}
    \caption{Posterior probability distributions of parameters for AT\,2018bwo. The model has four free parameters: Tstar, Tdust, tau\_{100um}, and sil, and r$_{in}$, L, M$_{\rm{dust}}$ are derived parameters. Note that \texttt{tau\_{100um}} represents the dust optical depth at 100\,\um.}
    \label{fig:corner_at2018bwo}
\end{figure}

\begin{figure}
    \centering
    \includegraphics[width=0.8\textwidth]{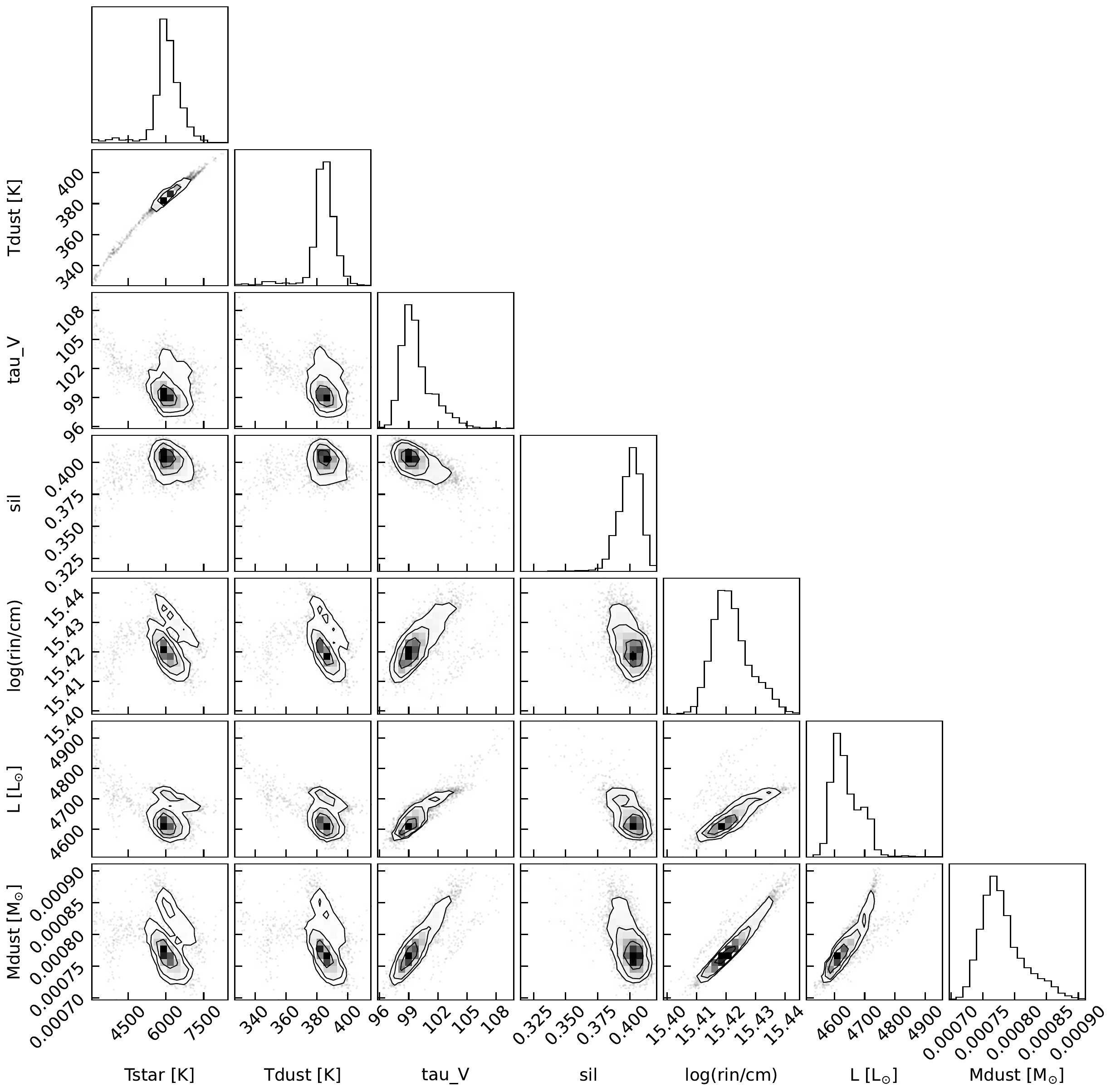}
    \caption{Posterior probability distributions of parameters for M31-LRN-2015. The model has four free parameters: Tstar, Tdust, tau\_{V}, and sil, and r$_{in}$, L, M$_{\rm{dust}}$ are derived parameters.}
    \label{fig:corner_m31lrn2015}
\end{figure}
\end{appendix}
\bibliography{myreferences}
\end{document}